\documentclass{pasa}
\usepackage{subfig}
\usepackage{graphicx}
\usepackage{bm}
\usepackage[authoryear]{natbib}
\bibpunct{(}{)}{;}{a}{}{,}
\title[Initial conditions in N-body simulations of debris discs]{The impact of initial conditions in N-body simulations of debris discs}
\author[E. Thilliez et al.]{E. Thilliez$^{1}$\thanks{e-mail: ethilliez@astro.swin.edu.au} \and S. T. Maddison$^1$\\
\affil{$^1$Centre for Astrophysics and Supercomputing, Swinburne University of Technology, Hawthorn, VIC 3122, Australia}}%

\jid{PASA}
\doi{10.1017/pas.\the\year.xxx}
\jyear{\the\year}

\begin{document}
\begin{abstract}
Numerical simulations are a crucial tool to understand the relationship between debris discs and planetary companions. As debris disc observations are now reaching unprecedented levels of precision over a wide range of wavelengths, an appropriate level of accuracy and consistency is required in numerical simulations to confidently interpret this new generation of observations. However, simulations throughout the literature  have been conducted with various initial conditions often with little or no justification. %
In this paper, we aim to study the dependence on the initial conditions of N-body simulations modelling the interaction between a massive and eccentric planet on an exterior debris disc. %
To achieve this, we first classify three broad approaches used in the literature and provide some physical context for when each category should be used. We then run a series of N-body simulations, that include radiation forces acting on small grains, with varying initial conditions across the three categories.  We test the influence of the initial parent body belt width, eccentricity, and alignment with the planet on the resulting debris disc structure and compare the final peak emission location, disc width and offset of synthetic disc images produced with a radiative transfer code. We also track the evolution of the forced eccentricity of the dust grains induced by the planet, as well as resonance dust trapping. %
We find that an initially broad parent body belt always results in a broader debris disc than an initially narrow parent body belt. While simulations with a parent body belt with low initial eccentricity ($e \sim 0$) and high initial eccentricity ($0<e<0.3$) resulted in similar broad discs,  we find that purely secular forced initial conditions, where the initial disc eccentricity is set to the forced value and the disc is aligned with the planet, always result in a narrower disc. %
We conclude that broad debris discs can be modelled by using either a dynamically cold or dynamically warm parent belt, while in contrast eccentric narrow debris rings are reproduced using a secularly forced parent body belt.
\end{abstract}
\begin{keywords}
planet-disc interactions - circumstellar matter - methods: numerical
\end{keywords}
\maketitle

\section{Introduction}
Recent images of the  $\sim40$ currently known resolved debris discs show both radial and azimuthal structures, such as gaps \citep{2009ApJ...705..314S}, eccentric rings (Kalas et al., 2005) and warps \citep{2000ApJ...539..435H}. Although asymmetries can result from non-planet interaction \citep{2013Natur.499..184L}, these configurations likely result from planetary companions shaping the disc by their gravitational influence \citep{2002ApJ...578L.149Q,2005ApJ...625..398D,2012A&A...537A..65T}. 

N-body simulations are a common tool to model the dynamical evolution of debris discs by following the trajectories of individual grains in the disc. By numerically integrating the Kepler and perturbation equations, these N-body codes allow us to study the dynamical perturbation of planetary companions on the structure of debris discs. The standard simulation scenario is as follows: assuming the disc is in a collisional steady-state, a parent body belt of asteroid or comet-like bodies will create a collisional cascade with a constant mass loss rate. The grains, modelled by massless test particles, initially have similar orbital parameters to their parent bodies, but because the grains are sensitive to radiation forces as well as the gravitational perturbation of any planetary companions, the particles will evolve on different orbits. 

Together with continued improvements in observational techniques, analytical perturbation theory for debris discs has been developed to describe the evolution of the disc under the sculpting influence of a planet.  
For example, an eccentric or inclined planet induces secular perturbations in the disc by imposing an eccentricity or inclination on the grains and thus creating brightness asymmetries or eccentric rings \citep{1999ApJ...527..918W}. This theory was used to explain the brightness asymmetry in the debris disc of HR 4796A. More recently, \cite{2012A&A...542A..18L} found the planet $\beta$ Pic b to be responsible for the inclined component of the disc. \cite{2012AJ....144...45K} noticed that the star HD 202628 is displaced from its debris discs centre by 20~AU and strongly suspect a distant planet to be responsible for the offset.
Resonant interactions have also been analytically studied in conjunction with numerical simulations \citep{2003ApJ...588.1110K,2003ApJ...598.1321W}, although theoretical results and simulations can be difficult to reconciled, mainly because they are often built on different initial assumptions.

The dynamical evolution of a debris disc under the gravitational influence of a planet is a complex mixture of (i) grain collisions of different sizes which continuously reshape the grain size distribution, (ii) stellar radiation forces affecting the orbits of each grain size population differently and (iii) a mixture of resonant and secular perturbations arising from the planetary companion. While pure N-body simulations only model the interaction of limited grain populations without any size distribution evolution or collisions \footnote{Although we note recent progress in incorporating grain-grain collisions and its feedback on the dynamics with the LIDT-DD \citep{2013A&A...558A.121K} and SMACK codes \citep{2013ApJ...777..144N}.}, analytical predictions for debris discs are, in addition to the N-body limitations on sample and distribution size, based on the generalisation of a restricted three body single interaction case (resonant or secular) limited at the first or second order. Therefore, numerical N-body simulations represent the best choice to accurately model debris discs interacting with planets, while still assuming a range of simplifications such as a limited grain sample with no size distribution evolution or collisions. With improvements in the resolution and sensitivity of the instruments at both short and long wavelengths (e.g. \textit{James Webb Space Telescope} (JWST) and \textit{ Atacama Large Millimeter/submillimeter Array} (ALMA) respectively), we must ensure that the best comparison between debris disc models and observations can be made. One fundamental step is to check that any numerical simulations will produce consistent results with few (or no) dependancies to the initial parameters.
 
The aim of this paper is to study the impact of changing the initial conditions in a simulation where a debris disc is interacting with an eccentric and massive planet on a fixed orbit. We run a suite of simulations over a broad range of initial conditions covering the three disc types that we classify from the literature. We then quantify the impact of the initial conditions on the resulting disc structure (specifically the disc offset, brightness profile, disc mean radius and width) at different stages of the simulation. We also track the resonant and secular evolution of the grains in the disc. 

The paper is organised as follows: in Section 2 we discuss the different types of initial conditions encountered in the literature and provide some physical context for each class, then in Section 3 we describe our methodology for the simulations and the creation of synthetic images, as well as how we extract the final disc parameters from the synthetic images. We compare the results of the different simulations in Section 4, and discuss our results and conclusions in Section 5. 

\section{Classifying disc initial conditions}

\begin{figure}[h]
\centering
\includegraphics[width=0.5\textwidth]{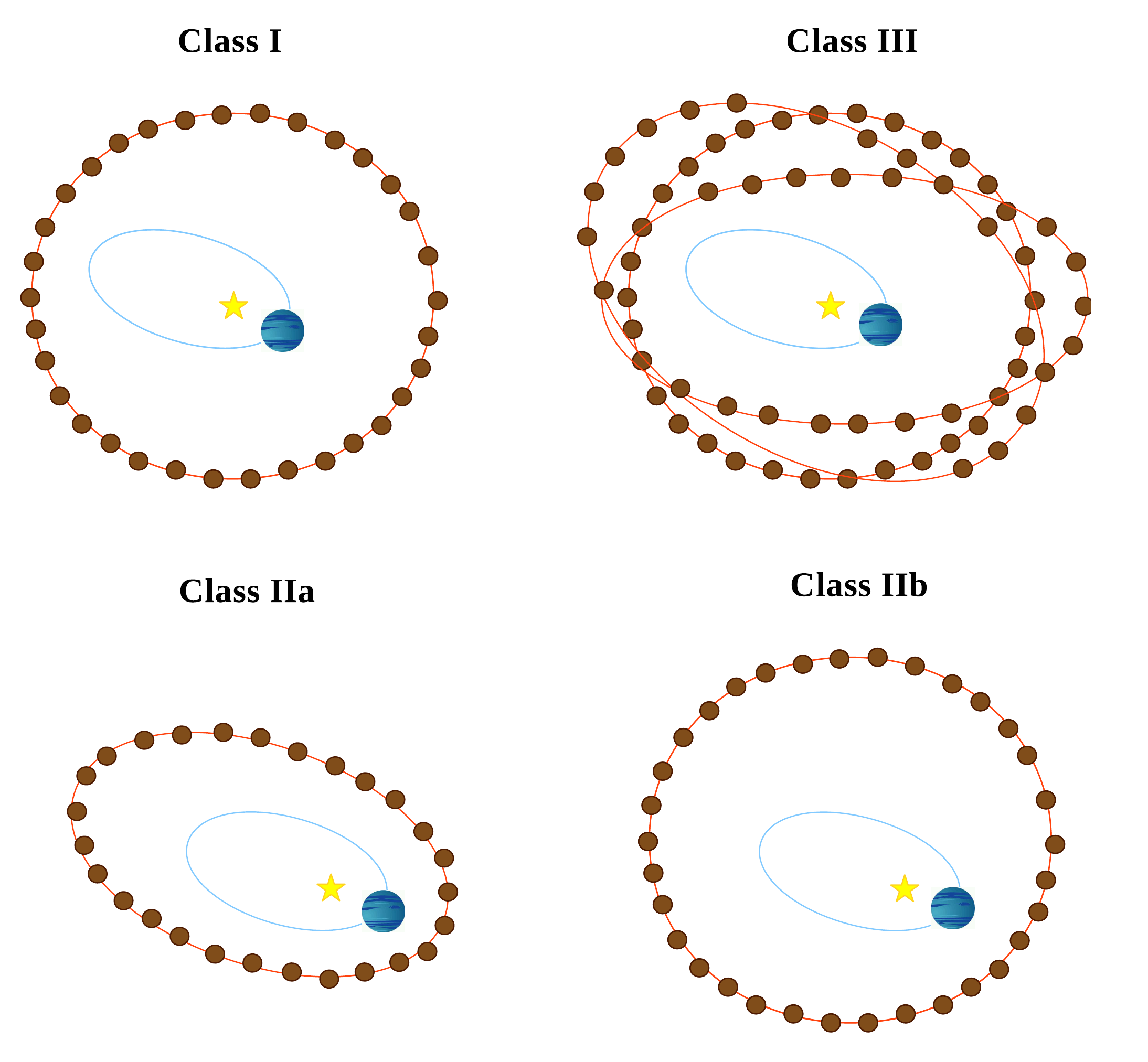} 
\caption{Different classes of initial conditions for numerical simulations used throughout the literature. Class I, dynamically cold discs; Class II, forced discs; Class III, dynamically warm discs.}
\label{fig1}
\end{figure}

Depending on the specific goals of any study, a range of different initial conditions are used in the literature and often with little justification. We have divided these initial conditions into three broad classes (see Figure \ref{fig1}), and here provide some physical context for each class.

\begin{itemize}
\item Class I: the dynamically cold disc. This scenario physically follows from the protoplanetary disc phase, where the eccentricity of a low mass embedded planet, $e_{p}$, is damped by the gaseous disc \citep{1993ARA&A..31..129L,2007A&A...473..329C}. Therefore the disc has a low eccentricity excitation and remains quasi-circular with $e \sim 0.0$. The planet's eccentricity can increase during the planet-planet scattering phase \citep{2003ApJ...584..465C} or later merging events. At the beginning of these simulations, the planet is assumed to obtain its mass and eccentricity instantaneously, without perturbing the circular disc \citep{2005A&A...440..937W}. Spiral features can be temporarily created due to planetesimals at different distances having different precession periods \citep{2005A&A...440..937W} and over time the disc eccentricity will increase to a final value set by the planet's eccentricity \citep{2014A&A...563A..72F}. 
This initial configuration was used by \cite{2014MNRAS.443.2541P} in their numerical study of a debris disc interacting with a highly eccentric planet. In the case of a coherent disc with initial $e< 0.08$, this interaction resulted in a disc apse aligned with the planet, with the eccentricity of the outermost particles of the disc similar to the expected forced eccentricity after a few secular times. A similar initial configuration was used by \cite{2014A&A...561A..43B} when trying to reproduce the debris disc of Fomalhaut: using a initially cold disc ($e<0.05$), the interaction between the disc and a super-Earth planet with eccentricity $e_{p}=0.8$ led temporarily ($t< 20$~Myr) to a debris disc with the observed eccentricity. However the disc was not apse aligned with the planet but rather appeared rotated by $70^{\circ}$.  
\item Class IIa: the forced, apse aligned disc. This initial configuration relies on analytical predictions and assumes that secular perturbations have sculpted an eccentric debris disc and apse aligned it with the planet. This means that the major axes of the disc and planet are aligned. Specifically it forces the disc's vector eccentricity, $\textbf{e}$, to vary on a circle of radius $e_{free}$ around the forced value $e_{forced}$, with the direction of $e_{forced}$ and $e_{free}$ respectively set by the forced and free orientation of the pericentre, $\overline{\omega}_{forced}$ and $\overline{\omega}_{free}$. Therefore, by initially setting the apse alignment $\overline{\omega}=\overline{\omega}_{forced}$ and the eccentricity equal to the forced value, $e=e_{forced}$, the eccentricity of the disc particles are expected to be purely forced with no free component ($e_{free}=0$). While this may be acceptable for massive and eccentric planets \citep{2005ApJ...625..398D}, in practice the planet-disc alignment is not always apparent in numerical simulations \citep{2014A&A...561A..43B}. Moreover the value of the predicted forced eccentricity is a function of semi-major axis \citep{1999ApJ...527..918W}, and as noted by \cite{2014A&A...563A..72F}, setting a value of the initial forced eccentricity for a broad debris disc can be problematic. However, this initial configuration has been used by \cite{2009ApJ...693..734C} while investigating if the inner sharp edge of the Fomalhaut disc could result from planetary forcing. \cite{2006MNRAS.373.1245Q} investigated the chaotic zone surrounding eccentric planets and argued that this scenario could take place in high collisional discs, where inelastic collisions damp the free eccentricity of the parent body belt and force their alignment and eccentricity by secular interaction with a planet. 
\item Class IIb: the forced, quasi circular apse aligned disc. Again, this initial configuration assumes that the disc is apse aligned with the planet but remains quasi circular. With $e \sim 0$ and $\overline{\omega}=\overline{\omega}_{forced}$, this configuration is expected to produce a broad disc with particle eccentricity ranging from $0 <e <2e_{forced}$ \citep{2009ApJ...693..734C}. \cite{2014MNRAS.438.3577T} started simulations with the forced planet alignment with a quasi circular disc ($e < 0.01$) and reproduced the ring of Fomalhaut for $t<0.4$ Myr (although the disc apse alignment appeared rotated by $90^{\circ}$ to the planet), before particles moved to high eccentricity orbits. \cite{2009ApJ...693..734C} ran a second set of simulations using the circular forced disc as initial condition for the Fomalhaut disc and found the resulting disc structure too broad to match the observed narrow ring after 100 Myr.
\item Class III: the dynamically warm disc. In the case of a protoplanetary disc harbouring a massive planet,
the disc-planet interaction can result in a gap opening \citep{2013A&A...555A.124B}, and the orbit of the planet can become eccentric \citep{2001A&A...366..263P}. If the gap is wide enough, not only is the damping of the planet's eccentricity reduced \citep{2003ApJ...585.1024G}, but the outer Lindblad resonance can also excite the disc eccentricity up to $e \sim 0.3$ \citep{2006A&A...447..369K}. This scenario was used by \cite{2005ApJ...625..398D} for a disc with initial $0< e< 0.3$ to reproduce the Vega clumps with a 3 Jupiter mass planet of $e_{p}=0.1$.  
\end{itemize}

In addition, some studies in the literature use a fourth configuration whereby the parent body belt is initially reduced to the infinitely narrow belt (rather than a broader belt of a few tens of AU). To study the observed debris disc width in scattered light with potential companion parameters, \cite{2014ApJ...780...65R} used an initially thin and forced parent body belt in order to avoid any degeneracy between the initial disc width and the planetary mass needed to shape the disc. A similar configuration was used by \cite{2002ApJ...569L.115W} while modelling the resonant clumps around Vega.

\section{Methods}
 In the literature, we find different types of initial disc conditions that can broadly be classified into three classes of initial conditions discussed in Section 2. Each class differs by having a different disc-planet orientation, as well as different initial eccentricity and width of their parent body belt from which the grains are released. The disc-planet orientation can be either free (without any constraints) or forced (or apse aligned, meaning the direction of the disc pericentre is forced to be aligned with the direction with the planet pericentre by secular perturbations). The grain's eccentricity vector, which points towards the periapsis with a magnitude corresponding to the elliptical orbit eccentricity, can have two components: a free component corresponding to the grain's intrinsic orbital eccentricity and a forced component that the grain inherits from the secular forcing of the planetary companion. Some initial conditions have a disc with a forced eccentricity, while others have either low free eccentricity or a high free eccentricity. 

In this work we use the three broad disc classifications described above to study the effects of different initial conditions on the structure of debris discs. In this section we first describe the numerical code we use for our simulations and the planet and disc parameters used in our models. We also describe the code used to produce of synthetic images in order to study the resulting structure of the discs (so that these can be compared with observations) and detail the method used to determine disc parameters for the synthetic models. 

\subsection{Simulations}
We use a modified version of the N-body symplectic integrator \textit{SWIFT} \citep{1994Icar..108...18L} in which we have incorporated radiation forces. Small grains in the disc are sensitive to the stellar radiation forces such as radiation pressure, Poynting Robertson (PR) drag \citep{1979Icar...40....1B} and stellar wind drag \citep{1982A&A...107...97M}. The radiation pressure is described by the fraction, $\beta$, of the gravitational force between the central star and the grain, where $\beta$ decreases linearly with grain size:
\begin{equation}
\beta=0.577 \frac{L_{\ast}}{L_{\odot}} \left(\frac{\rho}{\rm{g/cm}^{3}}\right)^{-1} \left(\frac{s}{\mu\rm{m}}\right)^{-1} \left(\frac{M_{*}}{\rm{M}_{\odot}}\right)^{-1},
\label{eq:1}
\end{equation}
where $L_{\ast}$ and $M_{\ast}$  are the stellar luminosity and mass, and $\rho$ and $s$ are the intrinsic grain density and size. 
 The total acceleration felt by the grains is given by:
\begin{equation}
\frac{d^{2} \bm{r}}{dt^{2}}=F_{\rm grav}\left(\beta \bm{r} -\frac{\beta(1+sw)}{c}(v_{r}\bm{r}+\bm{v})\right),
\label{eq:2}
\end{equation}
where \textbf{\textit{r}}, \textbf{\textit{v}} are the position and velocity vectors and $sw$ is the ratio of solar wind drag to radiation pressure. We use a stellar mass and luminosity equal to $M_{\odot}$ and $L_{\odot}$ in all simulations, and model grains of size $s=13$~$ \mu$m with density $\rho=3.5$~g/cm$^{3}$, so that the ratio of radiation pressure to the gravitational force is $\beta=0.02$.  Since silicate is a typical dust composition at the radial distance corresponding to the disc location in a solar-type environment ~\citep{2015ApJ...798...87M}, we set $sw=0.05$.

We use a 2 Jupiter mass planet with an orbital eccentricity $e_{p}=0.3$ at a semi-major axis of $a_{p}=30$ AU, and the planet starts at its pericentre with $\omega_{p}=0^{\circ}$. This configuration was chosen to enhance the disc features resulting from the disc-planet interaction: a shorter secular timescale and a higher dust trapping efficiency is expected from a massive planet, while the secular induced eccentricity is directly proportional to the planet eccentricity. We employ this unique planetary configuration for all classes of initial conditions, despite reporting in Section 2 that a 2 Jupiter mass planet is below the minimal value for a planet to open a gap ($> 5$ Jupiter masses, Bitsch et al., 2013) in the protoplanetary disc. The creation of this gap by a massive planet is normally required for the initial condition of dynamically hot disc (Class III) to occur, while dynamically cold discs (Class I) are expected to be found in system with a circular planet. The chosen initial planetary conditions are therefore a trade-off between the conditions expected by each class. We did, however, run additional simulations using a 2 Jupiter mass planet at 30~AU on a quasi circular orbit ($e_{p}=0.03$) to check that our choice of initial conditions had little impact on our results. Further discussion is provided in Section 5.

 We model the debris disc using massless test particles representing an ensemble of dust grains. Because grains can be removed by either PR drag or radiation pressure, and the disc is assumed to be in a collisional steady state, keeping track of all the particles in the system is numerically challenging. We model the disc using 1200 massless test particles and record their positions and velocities every 7.25 planetary revolutions, 7.25$P_{p}$, and then stack the recorded frames to obtain the final particles distribution. N-body simulations are collisionless and therefore suitable to model low-collisional discs. While in a real system, grain-grain collisions could potentially inhibit some planet induced structures in the disc \citep{2009ApJ...707..543S}, our study here purely aims to investigate the impact of initial conditions in simulation outcomes. Therefore collisions are not modeled dynamically but rather approximate: stacking the particles distribution at regular intervals not only increases the total number of particles, but also allows us to approximate the replenishment of small grains created from planetesimal collisions within the parent body belt as expected in a steady state system. 

In order to cover the three classes of initial conditions, we create a range of initial disc configurations by varying three parameters: (1) the eccentricity range, (2) the semi-majox axis range (and hence the disc width) and (3) the longitude of pericenter of the particles. We model each class of discs with  a broad parent body belt located between $45<a<80$~AU  and a narrow belt located at $67.5<a<67.6$ AU to avoid setting the parent body belt in the proximity to any resonances (see Table ~\ref{Table3}). For the disc eccentricity, we model a quasi circular disc with $0<e<0.04$, a disc with forced eccentricity, $e_{forced}$, and with random eccentricity $0<e<0.3$. Using the approximation of \cite{2014MNRAS.443.2541P}, we estimate the forced eccentricity at 67.5 AU for our planet to be $e_{forced}=0.17$. For the longitude of periaspsis, the discs have either a random orientation or a forced apse alignment with the planet, $\overline{\omega}=\overline{\omega}_{p}$. See Table \ref{Table1} for a summary and Table \ref{Table2} for the full suite of initial configurations. 
\begin{table}
\centering
\caption{Parameter range used in each initial condition class.}
\label{Table1}
\begin{tabular}{ccc}
\hline \hline
Parameter & range & definition \\
\hline
$\overline{\omega}$ & $0<\overline{\omega}<2\pi$& random\\
$\overline{\omega}$ & $\overline{\omega}=\omega_{p}$& forced\\
$e$ & $0<e<0.04$ & small \\
$e$ & $e=e_{forced}$ & forced \\
$e$ & $0<e<0.3$ & random \\
$a$ & $45<a<80$~AU & broad \\
$a$ & $67.5<a<67.6$~AU & narrow \\
\hline
\hline
\end{tabular}
\end{table}

Following \cite{1999ApJ...527..918W}, the secular precession timescale is expressed by $t_{sec}=2\pi/At_{year}$ with $t_{year}$ the number of seconds in a year, and $A$ being the precession rate which is a function of the planet $a_{p}$ \& $m_{p}$ and the disc semi-major axis, $a$, and mean motion, $n$. We estimate the secular precession timescale for a grain at $a=80~$AU (the outer edge of our broad disc) interacting with the 2 Jupiter mass at $a_{p}=30~$AU to be $t_{sec}=2.7\times 10^{6}$ years. The PR drag timescale, defined as $t_{PR}=400a^{2}/\beta$, for clearing a grain initially orbiting at $a=67.5~$AU in this stellar environment is $t_{PR}=9.1\times 10^{7}$ years. Therefore we choose a simulation duration of $t_{sim}=27$~Myr, corresponding to 10 $t_{sec}$ \footnote{We note that the forced eccentricity, $e_{forced}$, and the secular timescale, $t_{sec}$, are both a function of the semi-major axis -- see Figure \ref{fig4}. We choose to calculate the secular timescale at the outer edge of the disc to account for the delay of secular interaction in the outer disc. We calculate the forced eccentricity value at 67 AU in the initial conditions in order to accurately model for the narrow parent body belt case.}, which is long enough for the secular and resonance interactions to occurs before being destroyed by PR drag.

\subsection{Synthetic images}
Since our aim is to study the consequences of changes in the initial conditions on resulting debris disc structures, we use the 3D Monte Carlo radiative transfer code \textit{MCFOST} \citep{2006A&A...459..797P} to produce synthetic images from our simulations. We make synthetic images corresponding to mid-infrared observations at $\lambda=24~\mu$m since this is the domain is where small grains of size $s=13~\mu$m dominate the emission. To estimate the emission, we stack all spatial distribution frames (effectively each data dump at $7.25~P_{p}$) from each simulation into a single stacked distribution, and convert the stacked distribution of test particles into a 3D density map. We assume that the total mass of the disc is one lunar mass and that this mass is contained in the total number of particles in the stacked distribution. The 3D density structure is binned onto an appropriate cylindrical grid and used as input for \textit{MCFOST}. We chose 90 radial bins between $20-120$~AU, corresponding to an average radial bin size of $\sim$ 1.1 AU. (For comparison, 1.1 AU represents the angular resolution of the future \textit{JWST/Mid-Infrared Instrument} at 24$~\mu$m for an object at a distance of 10 pc.) The software traces monochromatic photon packets isotropically escaping from the star and propagating throughout the disc and calculates the temperature structure of the disc. Synthetic images are then derived by tracing rays over the photons paths.

\begin{table*}[t]
\centering
\caption{Initial conditions and resulting disc structure for each model. The variables in the right side of the Table are the resulting: brightness profile peak location $r_{0}$, disc $\Delta r/r_{0}$ (with $\Delta r$ the profile FWHM), the disc offset $\delta$, the list of MMR occupied by test particles in the disc and the forced eccentricity $e_{forced}$ obtained by particles at 1 $t_{sec}$.}
\label{Table2}
\begin{tabular}{ccccc||ccccc}
\hline \hline
Model & Class &$\overline{\omega}$ & $e$ & $a$ & $r_{0}$ (AU) & $\Delta r/r_{0}$ & $\delta$ (AU) & MMR & $e_{forced}$ \\
\hline
1    & Class I & random & small & broad  & 49 & 0.45 & 14.9 & 5:2, 7:2, 4:1  & 0.16 \\ 
2    & Class I & random & small & narrow  & 52 & 0.24 & 11.8  & 5:2 & 0.19  \\ 
3    & Class IIa & forced & forced & broad & 48  & 0.3 & 15.0 & 5:2, 3:1, 4:1 & 0.14  \\ 
4    & Class IIa & forced & forced & narrow  & 62 & 0.2 & 6.6 & 3:1 & 0.13  \\ 
5    & Class IIb & forced & small & broad  & 50 & 0.47  & 15.1 & 5:2, 3:1, 4:1 & 0.17\\ 
6    & Class IIb & forced & small & narrow & 53 & 0.26 & 12.0 & 5:2, 3:1 & 0.19\\ 
7    & Class III & random & random & broad & 49 & 0.42 & 14.6 & 5:2, 7:2, 4:1 & 0.16\\
8    & Class III & random & random & narrow & 52 & 0.29  & 12.4  & 5:2  & 0.17 \\
\hline
\hline
\end{tabular}
\end{table*}

\begin{table}[h]
\centering
\caption{Location of the main mean motion resonances for a planet at 30~AU.}
\label{Table3}
\begin{tabular}{cccccc}
\hline \hline
MMR order & 7:3 & 5:2 & 3:1 & 7:2 & 4:1\\
\hline
$a$ (AU) &  52.5 & 55.0 & 62.0 & 69.0 & 75.5\\
\hline
\end{tabular}
\end{table}

\subsection{Parameter extraction}
We compare three different disc properties of the simulation results: the disc offset, the disc width ratio and the peak brightness location determined from the synthetic images, as well as the secular and resonance evolution of the disc. The section describes how each of these parameters is determined.
\subsubsection{Disc offset and surface brightness profile }
To determine the disc offset from the stellar position, $\delta$, we azimuthally cut the synthetic image into 120 bins and extract the brightest pixel in each bin. After extracting the coordinates of the 120 brightest pixels in each bin, we fit an ellipse to these pixels using a least-squares (deviation) method and obtain the offset coordinate compared to the location of the star via $\delta=\sqrt{x_{\rm off}^2+y_{\rm off}^2}$.

To determine the peak brightness location, $r_{0}$, and the disc width ratio, $\Delta r/r_{0}$, we radially cut the synthetic image into 90 rings and azimuthally average the surface brightness in each ring. We interpolate using a spline method between the 90 points to obtain a set of 500 data points of flux as a function of radius, $F(r)$. We then extract the peak brightness point, $r_{0}$, which corresponds to the maximal surface brightness. The disc width, $\Delta r$, is defined as the FWHM of the surface brightness profile. 

Both the disc offset and surface brightness profile are determined for 4 different epochs: 0.1, 1, 5 and 10 $t_{sec}$, in order to follow the evolution of the disc structure. 
\subsubsection{Dynamical evolution}
During the simulations, we record the semi-major axes occupancy in the disc at 4 different epochs: 0.1, 1, 5 and 10 $t_{sec}$. This allows us to follow the evolution of positions of the test particles and determine the main mean motion resonances (MMR) of each disc. Their locations are given in Table \ref{Table3}. 

To study the evolution of the disc-planet alignment, we first perform a visual check of the spatial distribution of the disc particles at different epochs: 0, 0.1, 0.5, 1 and 5 $t_{sec}$.  In order to track the secular evolution of the particle eccentricity, we plot the complex eccentricity map, ($e \cos \omega, e\sin \omega$), occupied by the test particles initially and at 3 different epochs 1, 5 and 10 $t_{sec}$. By using a K-mean cluster analysis, we determine the centre of the cluster of the test particles in the ($e \cos \omega, e\sin \omega$) map. Since secular perturbations by an eccentric planet force the eccentricity vector of the dust to evolve on a circle around the forced eccentricity value, this cluster centre corresponds to the forced eccentricity of the test particles imposed by the planet.

\subsection{Stability zone}
\begin{figure}
\centering
\includegraphics[trim=10mm 0cm 10mm 0mm,keepaspectratio=true,clip=true,width=90mm,height=75mm]{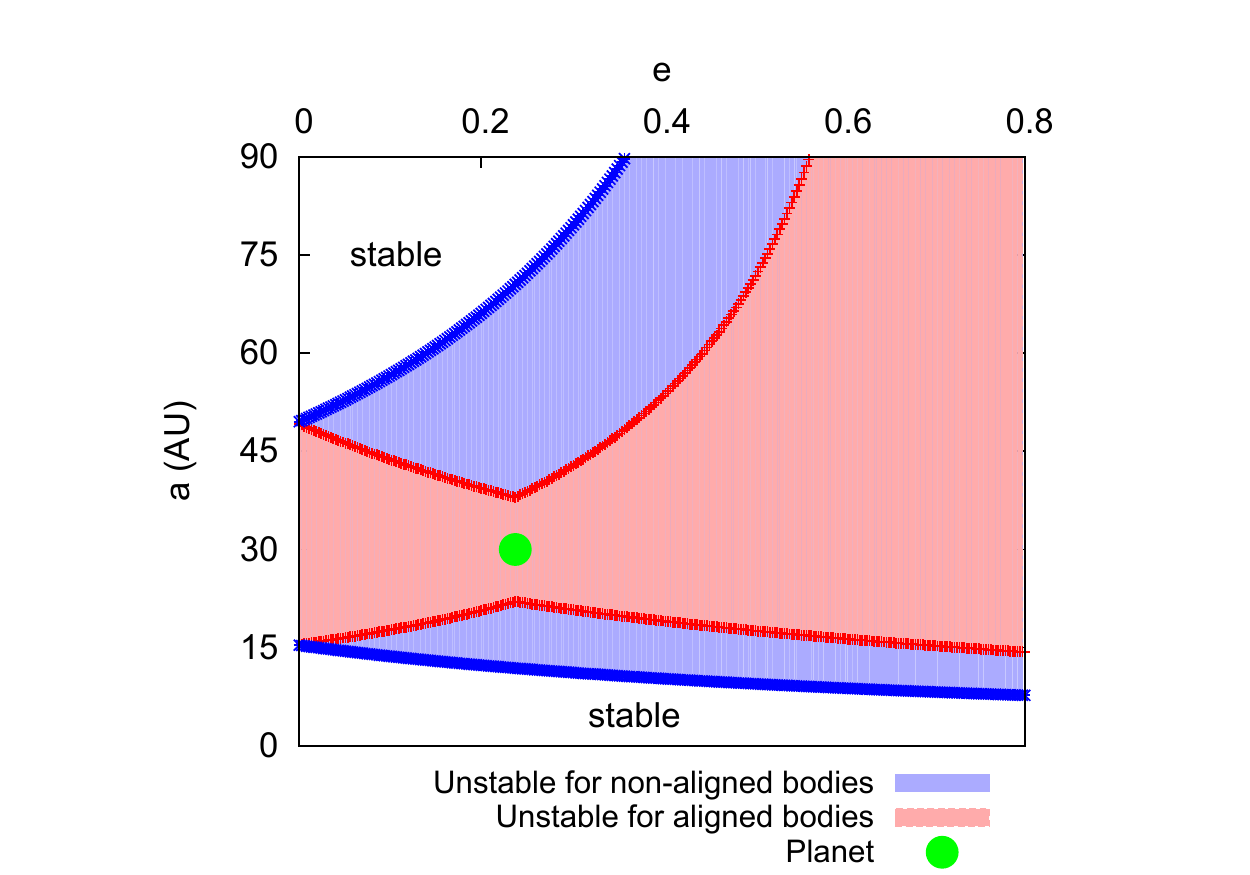} 
\caption{Theoretical stability map for our system. The green dot represents the location at 30~AU of the 2 Jupiter mass planet with $e_{p}=0.3$, while the lines delimit the extended orbit crossing regions between the massive planet and a potential companion the mass of Vesta when the asteroid is either apse aligned with the planet (red) or not (blue).}
\label{fig2} 
\end{figure}
To begin our analysis of how the dynamics of this planet-disc system responds to different initial conditions, we first predict the stable zones in the disc. We use the analytic criterion of \cite{2013MNRAS.436.3547G}, which constrains the location of the stable zone in the system by generalising the resonance overlap criterion from \cite{1980AJ.....85.1122W}. This criteria assesses the width of the region where, in the absence of any protective resonant mechanism, the proximity of two bodies will result in the chaotic diffusion of the eccentricity and semi-major axis evolution in one of the two orbits. The width of the region depends on the mass, $m_{p}$, eccentricity, $e_{p}$, and semi-major axis, $a_{p}$, of the massive planet, and on the mass and alignment of the encountering body.

To assess the location of the interior and exterior limit of the unstable zone around the giant planet, we assume that the second body has the mass of the asteroid Vesta. Therefore we are delimiting the chaotic zone around the giant and a Vesta like asteroid -- see Figure \ref{fig2}. Since we are modeling a planet interacting with an exterior debris disc, we only comment on the outer stability zone as we assume that grains interior to the planet's orbit will be quickly removed by the radiation forces. If the asteroid is not apse aligned with the planet, the asteroid should not survive in the inner $\sim$ 50 AU of the system unless trapped in a MMR. Moreover, the asteroid should keep a low to moderate eccentricity ranging from 0.2 at 60 AU to 0.45 at 80 AU in order to avoid close encounters with the planet. However, if the asteroid is apse aligned with the planet, the stable zone is much wider, allowing extreme eccentricities to be reached ($e \sim 0.6$). We can use this approximation to estimate the stability zone of our disc: to avoid being scattered by the planet, particles that are not apse aligned with the planet should avoid semi-major axes interior to 45-50 AU (unless trapped in MMR) and keep low to moderate eccentricities throughout the disc. 

Given the time for a grain at 65 AU to migrate to the inner unstable region at 45 AU is $t_{PR} \sim 8\times 10^{6}$, we can therefore predict that the PR drag will be a crucial factor in grain survival.
\begin{figure*}
\begin{center}  
\includegraphics[trim=0mm 0mm 24mm 24mm,keepaspectratio=true,clip=true,width=1.05\textwidth]{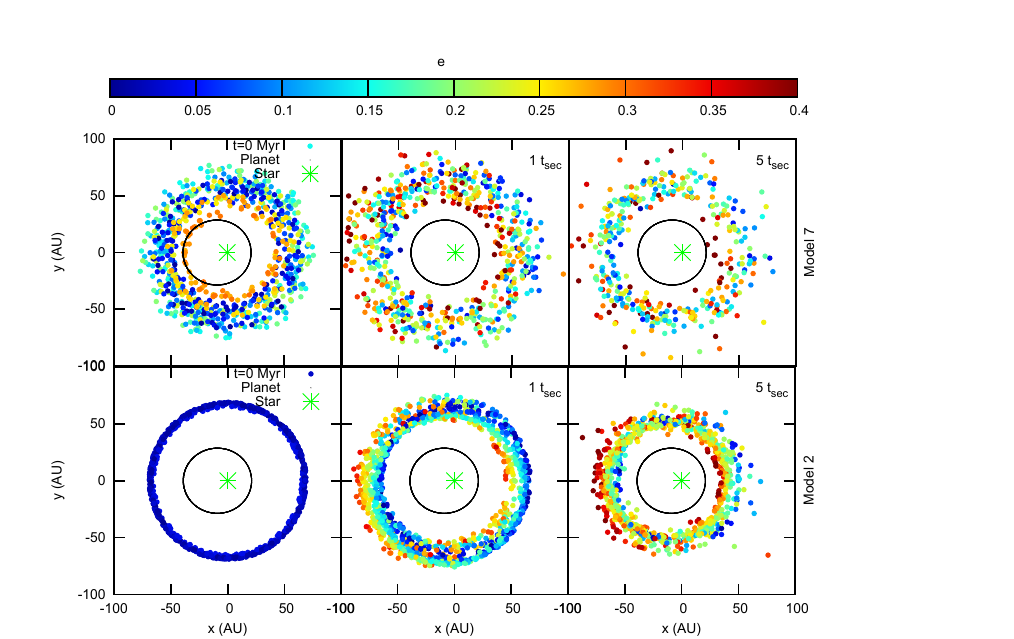}
\caption{Particle distribution evolution over three different epochs: 0, 1 and 5 $t_{sec}$ for initially non-aligned discs. The bottom row shows model 2 (Class I), corresponding to a disc with an initially narrow parent body belt, and the top row shows model 7 (Class III), corresponding to a disc with an initially broad parent body belt. The colorbar is the eccentricity of the particles. Both discs broaden after 0.5 $t_{sec}$ as the particle eccentricity increases and apse align with the planet within 1 $t_{sec}$. Similar conclusions are derived for any simulations with initial conditions from Class I or III.}
\label{fig3}  
\end{center}
\end{figure*}

\section{Results}
For the 8 different initial condition models we tested, we found two general outcomes. In the first case, if the simulation started with the initial conditions from Class I or III (corresponding to discs with a free orientation and either a  small or random eccentricity) or Class IIb (discs with a small eccentricity but apse aligned with the planet), the resulting structure is a broad disc apse aligned with the planet. A few differences appear when using a narrow parent body belt versus an extended belt, but the final disc structures are overall very similar. In the second case, simulations using the initial conditions from Class IIa  (discs with initial forced eccentricity and apse aligned with the planet), the resulting structure is a narrow disc apse aligned with the planet. Again, a few differences appear when using a narrow parent body belt versus an extended belt but overall the results are similar (these results are summarised in Figure \ref{fig13}). The resulting disc properties for each model are summarised in Table \ref{Table2}. In this section, we describe the dynamical evolution of these two outcomes. 

\begin{figure}
\centering
\includegraphics[trim=16mm 0cm 16mm 0mm,keepaspectratio=true,clip=true,width=80mm,height=55mm]{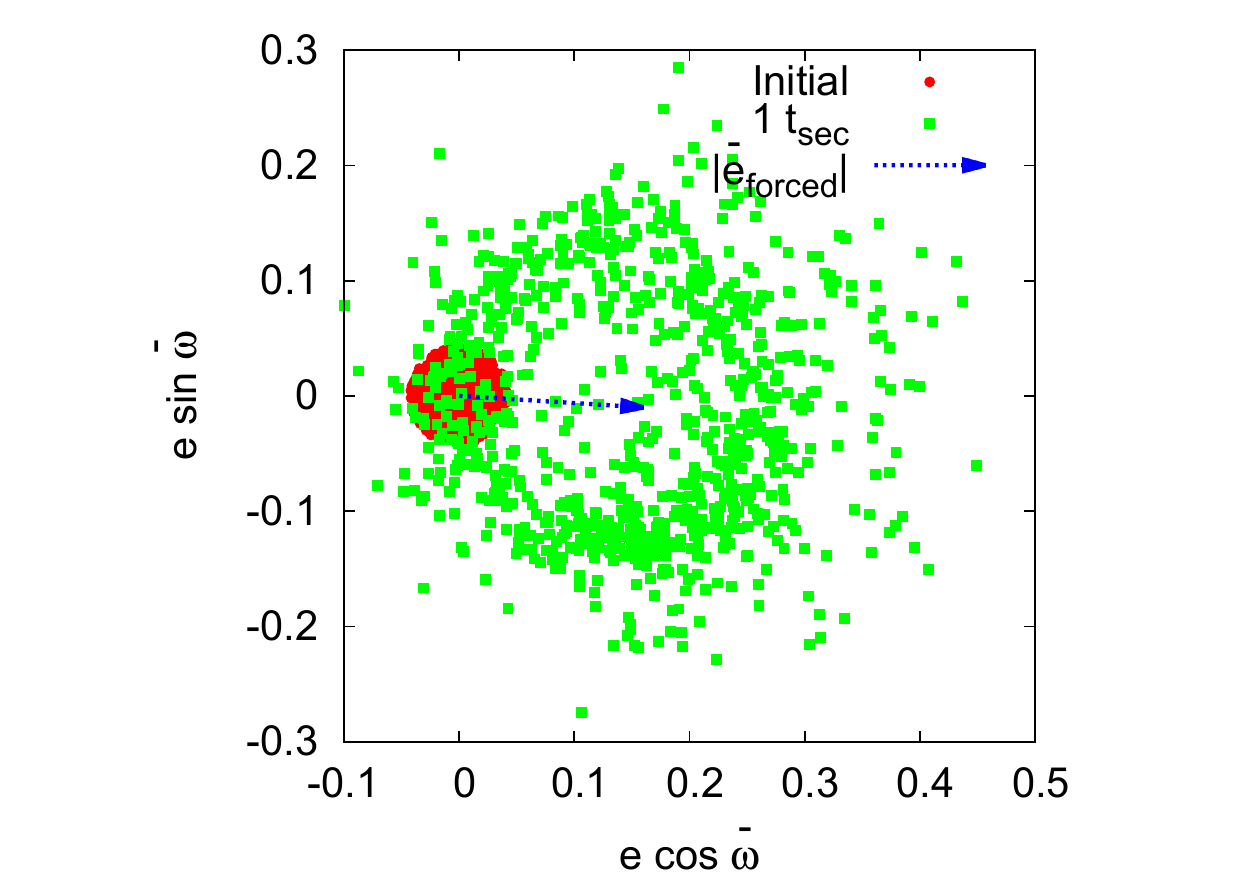} 
\caption{Complex eccentricity map ($e \cos \overline{\omega}, e\sin \overline{\omega}$) occupied by the particles initially (red) and after 1 $t_{sec}$ (green) for model 1. The complex eccentricities start to precess about the forced value in the direction of the forced pericentre ($\overline{\omega}_{p}=0$). The blue arrow points toward the particles forced eccentricity, $e_{forced} \sim 0.16$.}
\label{fig4} 
\end{figure}

\subsection{Producing broad discs}
Among the diversity of initial conditions, 6 of the 8 models resulted in a broad debris disc apse-aligned with the planet. Figure \ref{fig3} shows the evolution of the particle distribution for two very different initial disc setups: model 7 from Class III with a broad parent body belt (top) and model 2 from Class I with a narrow parent body belt (bottom). This illustrates that disc-planet apse alignment, as well as the disc broadening, occurs on a very short timescale.
\subsubsection{From a broad parent body belt}
Here we discuss simulations starting with an initially broad parent body belt and initial conditions from Class I, IIb, and III (corresponding to models 1, 5 and 7). In the simulations corresponding to these models, the disc particles undergo (i) radial inward migration due to radiation forces, (ii) a potential capture into near mean motion resonances, (iii) a secular forcing of their eccentricities with different forced values at different semi-major axes (which broadens the disc), and  (iv) if not initially the case (Class I \& III), the particles apse align with the planet. The overall resulting structure is a very broad disc apse aligned with the planet.

Figure \ref{fig4} provides a deeper understanding of the forcing on the eccentricities. In this case (model 1, Class I), the complex eccentricities were initially small ($0<e<0.04$), and then started to precess about the forced eccentricity (in the direction of the forced pericentre described by the blue arrow), which is induced by secular perturbations of the planet: since $\overline{\omega}_{p}=0$, the direction of the pericentre is the $x$ axis. Although particles are not initial forced ($e_{forced}=0$ at $t=0$), as soon as the simulation starts, the planet secularly forced the particles eccentricity. As a result, the particles complex eccentricities rapidly occupy a circle around the forced value with a radius equal to $e_{free}$ as seen in Figure \ref{fig4}. Because the initial complex eccentricities were close to zero, the particles' free eccentricity magnitude is equal to that of the forced value, and therefore the circle of the complex eccentricity encompasses the origin. With increasing eccentricity, the particles populate a wider region in the complex eccentricity phase space and as a result the disc gets broader.

An example of the evolution of the particle distribution when the planet and disc were not initially aligned (model 7, Class III) is shown in the top row of Figure \ref{fig3}. The disc aligns itself with the planet within 1 $t_{sec}$ and a broadening of the initial disc width can be seen after 0.5 $t_{sec}$. This disc broadening is due to test particle's eccentricity undergoing forcing from the planet as previously mentionned.\\

\begin{figure*}
\begin{center}               
  \includegraphics[width=170mm,height=117mm]{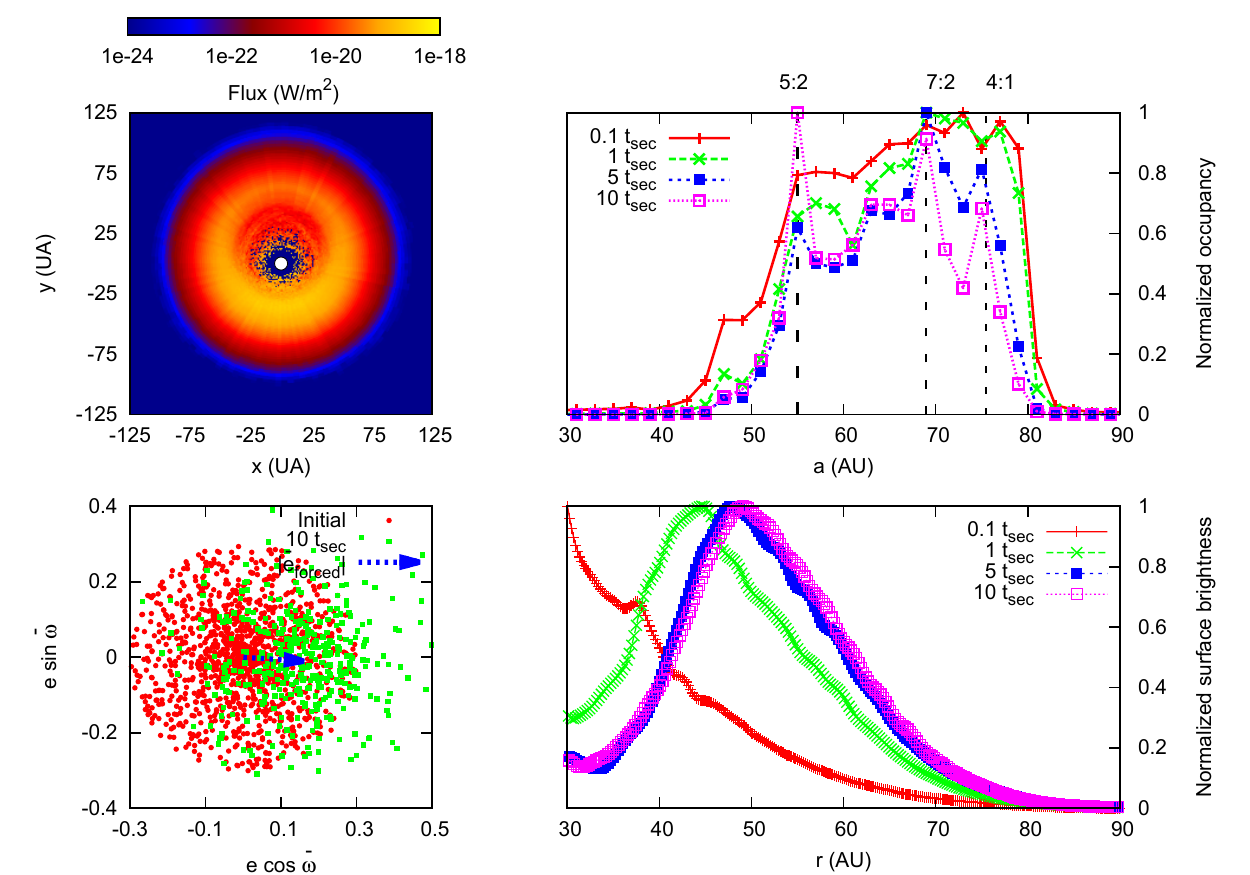}
  \caption{An initially broad parent body belt from Class III (model 7) results in a very broad disc. Top left: synthetic image at 24 $\mu$m which shows an eccentric ring with an offset of $\sim$ 15 AU from the stellar location. Top right: Normalized particle occupancy in the disc at 4 different epochs: 0.1, 1, 5 and 10 $t_{sec}$, highlighting the evolution of resonances in the disc. Bottom left: ($e \cos \overline{\omega}, e\sin \overline{\omega}$) complex eccentricity map with the blue arrow pointing toward the forced eccentricity, $e_{forced}$, occupied by the test particles initially (red) and after 10 $t_{sec}$ (green). Bottom right: Normalized radial surface brightness of the disc at different epochs: 0.1, 1, 5 and 10 $t_{sec}$.} 
\label{fig5}  
\end{center}
\end{figure*}

\begin{figure}
\centering
\includegraphics[width=80mm,height=53mm]{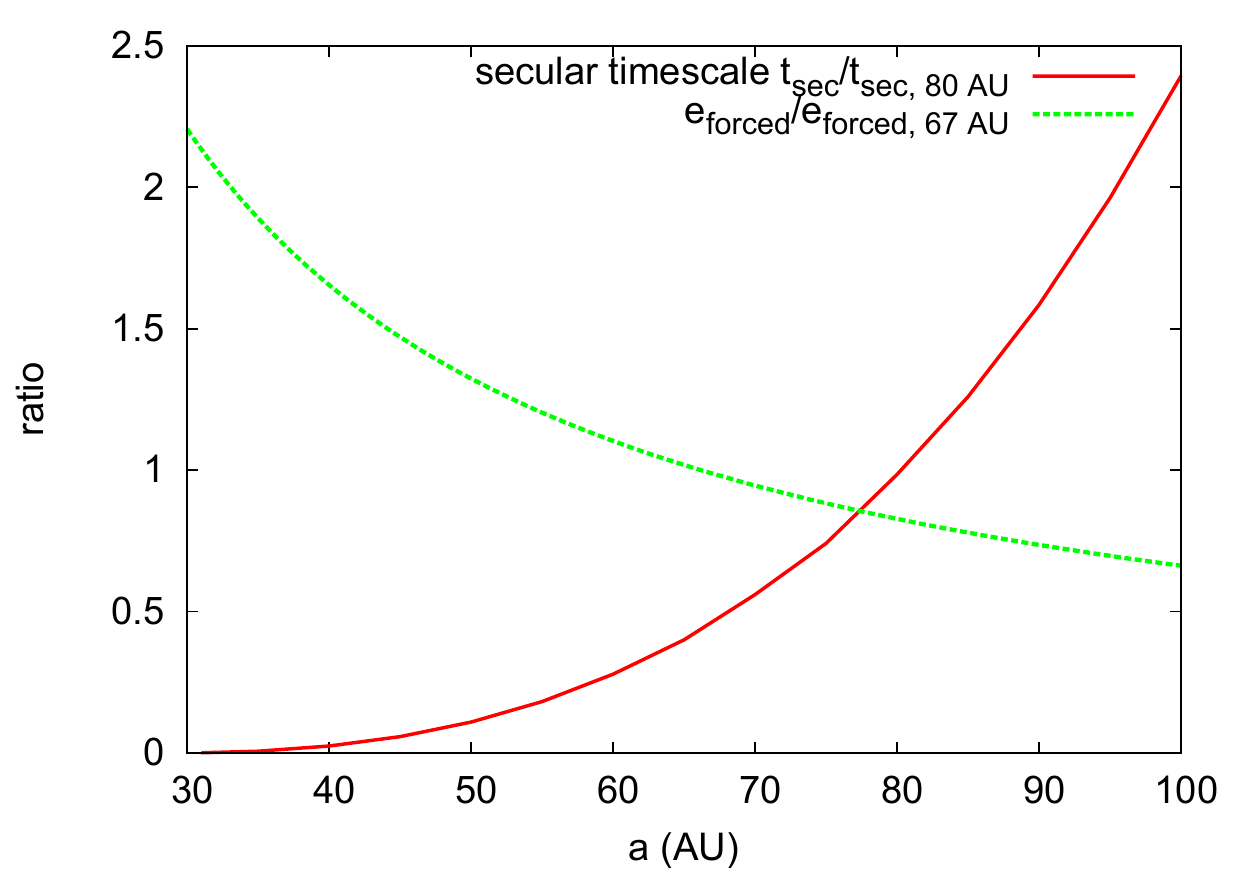} 
\caption{Approximations from the secular perturbations: ratio of the secular precession timescale to the secular timescale at 80 AU (red solid), and ratio of the forced eccentricity to the forced eccentricity at 67 AU (green dashed),  as a function of semi major-axis in a system with a 2 Jupiter mass planet located at 30 AU orbiting a solar star.}
\label{fig6} 
\end{figure}

Figure \ref{fig5} presents the final configuration of a typical broad disc outocme (in this case: model 7, Class III). Since the disc starts with a broad belt spanning $45<a<80$ AU, the disc inner particles are quickly affected by radiation forces, creating an inner component of hot dust seen in the synthetic image and the inner 40 AU of the surface brightness profile of Figure \ref{fig5}. The complex eccentricity map shows that particles roughly end up with $0<e<0.4$ after 10 $t_{sec}$. Because particles at various semi-major axes have different secular precession timescales and forced eccentricity values (see Figure \ref{fig6}), the complex map is a filled circle centred on the forced value, $e_{forced} \sim 0.16$, with a lot of scatter. The bottom panels in Figure \ref{fig5} show that the disc surface brightness and occupancy profiles reach their final state around 5 $t_{sec}$. Three MMRs -- 5:2, 7:2 and 4:1 -- are being populated, spanning $55 <a< 75$ AU. The final structure is a very broad disc with a width ratio of $\Delta r /r_{0} = 0.42$, peak brightness at $r_{0} = 49$~AU and disc offset $\delta = 14.6$~AU. The offset corresponds to a disc eccentricity of $e_{disc}=\delta/r_{0}=0.3$, similar to the planet eccentricity, $e_{pl}$. \\

The two remaining models in the group, models 1 and 5 of Table~\ref{Table2} which have an initially broad belt and initial conditions from Class I and IIb, have similar resulting disc parameters to model 7 within 5\% and a similar dynamical behaviour. 

\subsubsection{From a narrow parent body belt}
\begin{figure*}
\begin{center}               
\includegraphics[width=170mm,height=117mm]{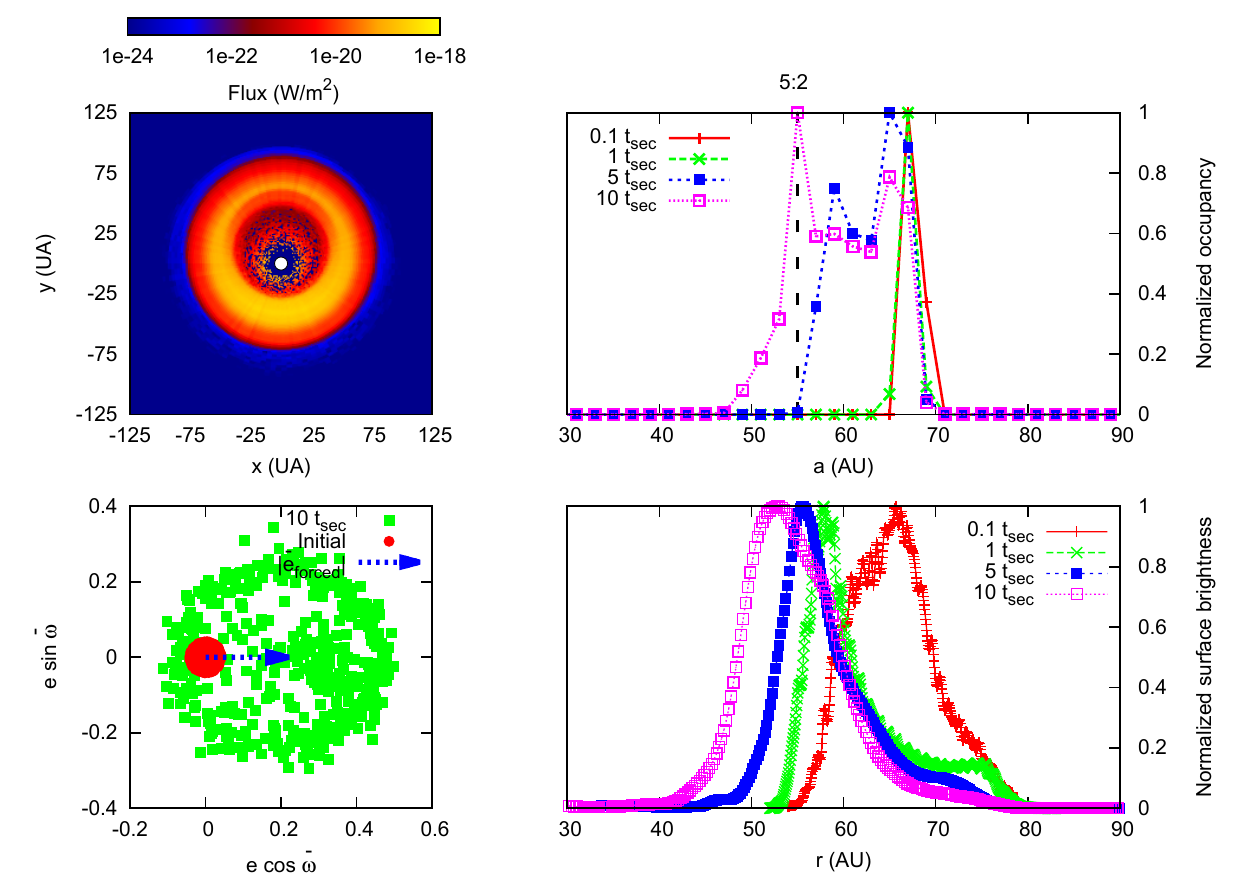}
  \caption{An initially narrow parent body belt from Class I (model 2) results in a broad disc. Top left: synthetic image at 24 $\mu$m which shows an eccentric ring with an offset of $\sim$ 13 AU from the stellar location. Top right: Normalized particle occupancy in the disc at different epochs: 0.1, 1, 5 and 10 $t_{sec}$, highlighting the evolution of resonances in the disc.  Bottom left:($e \cos \overline{\omega}, e\sin \overline{\omega}$) complex eccentricity map with the blue arrow pointing toward the forced eccentricity, $e_{forced}$, occupied by the test particles initially (red) and after 10 $t_{sec}$ (green). Bottom right: Normalized radial surface brightness of the disc at different epochs: 0.1, 1, 5 and 10 $t_{sec}$.} 
\label{fig7}  
\end{center}
\end{figure*}

Simulations starting with a narrow parent body belt and initial conditions from Class I, IIb, and III (corresponding to models 2, 6 and 8) exhibit a slighly different dynamical evolution. Figure~\ref{fig7} (corresponding to model 2) illustrates this evolution: the narrowness of the parent body belt (i) delays the time it takes for particles to populate the inner region $<$ 40 AU, (ii) prevents particles from reaching the outer MMRs such as the 4:1 (at 75 AU) or 7:2 (at 69 AU) within 10$~t_{sec}$, and (iii) makes the disc more sensitive to initial conditions because the belt confinement makes it sensitive to single events (like a strong MMR) that dominates the dynamics. The resulting disc structure is narrower (with a slighly smaller offset) than equivalent models with an intially broader parent body belt. Based on this result, we expect that such initial configurations will not be able to reproduce the very broad observed debris discs. 

The evolution of the particle distribution of an initially narrow parent body belt (model 2, Class I) is shown in the bottom row of Figure \ref{fig3}. It shows that, again, the disc rapidly broadens and becomes apse aligned with the planet within $1~t_{sec}$. 

To better understand the dynamics, the complex eccentricities map is presented in Figure \ref{fig7} top right: particles start with a complex eccentricity quasi null and experience eccentricity forcing, which lead their free eccentricities to increase in return, and therefore the complex eccentricities occupy a circle of radius $e_{free}$ around $e_{forced} \sim 0.15$ by $t=10~t_{sec}$. While the eccentricities range $0<e<0.4$ (like previous simulations with an initially broad belt), the main difference is that the inner region of the circle is less occupied with test particles (i.e. less scatter). This reflects that particles, initially located at a similar semi-major axis ($67.5<a<67.6$~AU), all move through the disc together, so they share similar $e_{forced}$ values, while particles from a broad parent body belt occupied a wider region of phase space due to their different forced eccentricities. 
Due to PR drag, the entire particle population migrates inwards, causing the brightness profile (bottom left panel of Figure \ref{fig7}) to move inward to $\sim$ 50 AU. The occupancy plot (Figure \ref{fig7} bottom right) shows that after an initial grain accumulation near the parent body belt, the interior 5:2 MMR (at 54 AU) becomes populated as particles migrate inwards. The final disc has $r_{0}=52$ AU, $\Delta r /r_{0} \sim 0.24$, and $\delta \sim 11.8$  AU which corresponds to $e_{disc}=0.23$.

The two remaining models in the group, models 6 and 8 of Table \ref{Table2} with an initially narrow belt with initial condition of Class IIb and III respectively, have similar disc parameters to model 2 within 8\%. While the peak brightness location of the simulations of Class I, IIb and III with either an initially broad or narrow parent body belt are consistent within 7\%, the disc width ratio is 40\% smaller in the narrow parent body belt case, with a smaller disc offset by 20\%.

Due to the secular forcing that causes particle eccentricities to evolve around $e_{forced}$ and eliminate orbits with extreme eccentricities ($e>0.4$) due to close encounters with the planet, it is not surprising that models from Class I with small initial eccentricities and Class III with random initial eccentricities result in a similar final disc structure. For the same reason, it is also not unexpected that models with initial conditions from Class IIb (which are expected to create a broad disc with $0 < e< 2 e_{forced} \sim 0.35$) results into similar structures to Class I and Class III.

\subsection{Producing narrow discs}
\subsubsection{From a broad parent body belt}
\begin{figure*}
\begin{center}               
  \subfloat{\includegraphics[trim=1mm 0cm 2mm 0mm,keepaspectratio=false,clip=true,width=65mm,height=48mm]{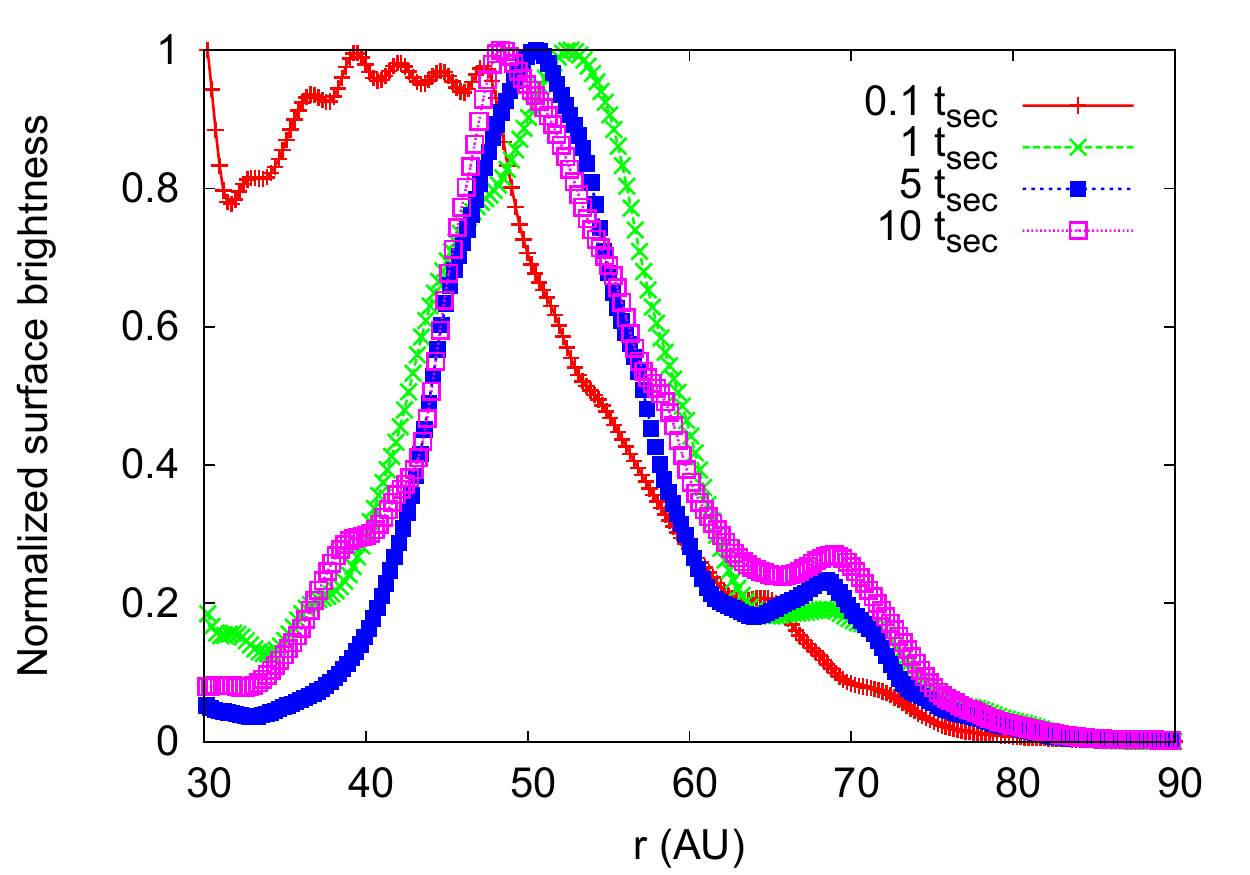}} 
  \subfloat{\includegraphics[trim=1mm 0cm 5mm 0mm,keepaspectratio=false,clip=true,width=65mm,height=48mm]{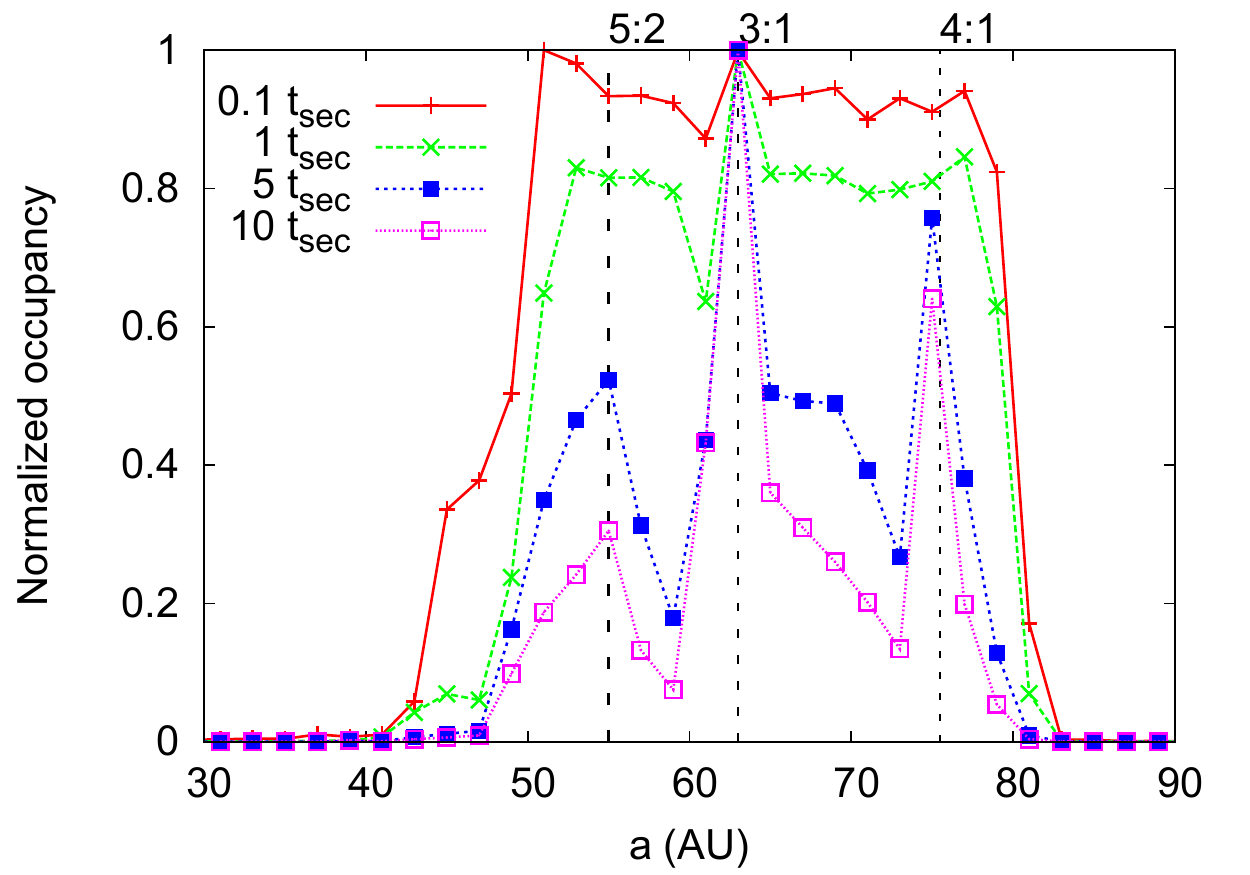}} 
  \subfloat{\includegraphics[trim=15mm 0cm 2mm 0mm,keepaspectratio=false,clip=true,width=65mm,height=48mm]{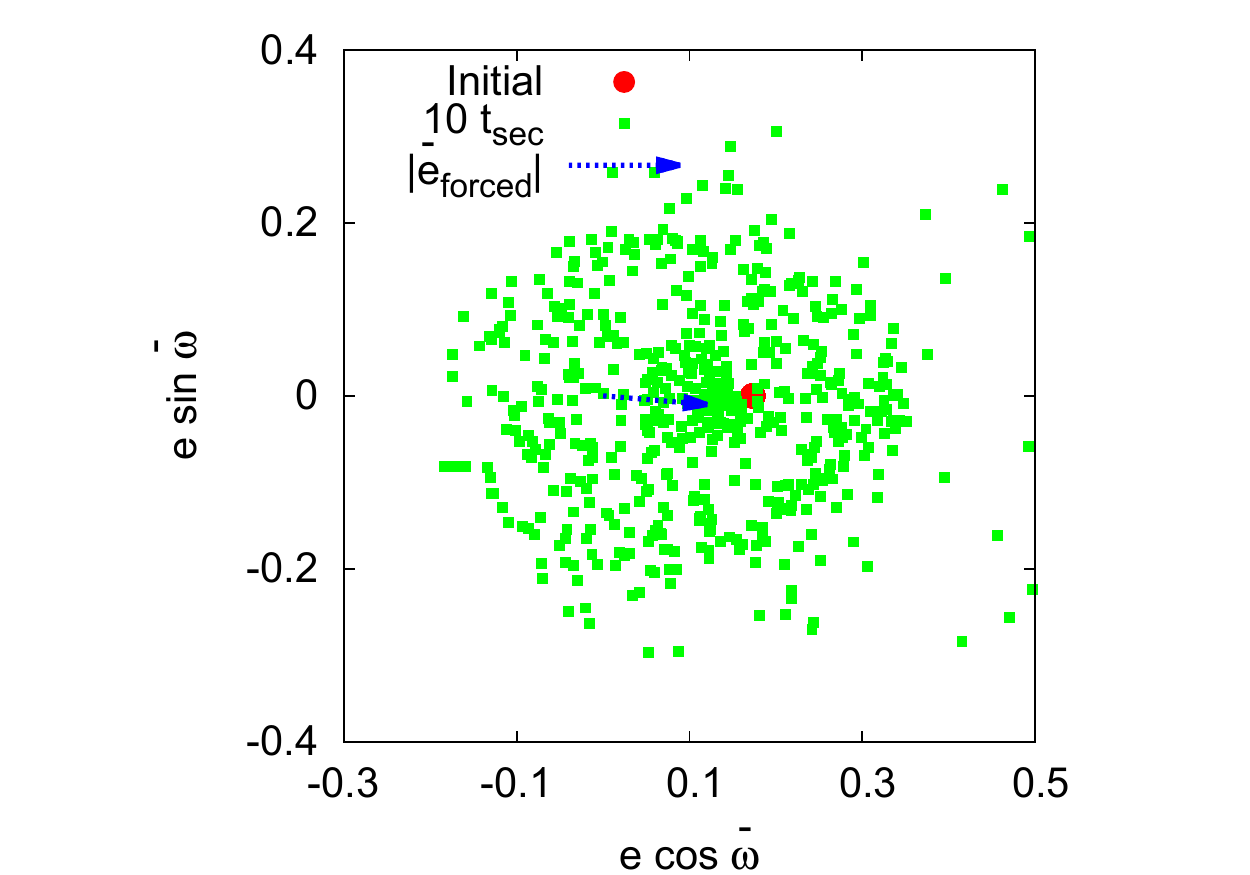}} 
  \caption{Results for model 3, corresponding to a disc with an initially broad parent body belt with initial conditions of Class IIa.} 
\label{fig8}  
\end{center}
\end{figure*}

Only one of our models starting with an initially broad parent body belt (model 3, Class IIa) resulted in a narrow disc - see Figure \ref{fig8}. In this scenario, where the disc initially has a forced eccentricity and is apse aligned with the planet, the particles undergo (i) a variation in their forced eccentricity values since the particles do not share the same semi-major axis, and (ii) are then trapped by the nearest MMR and therefore remain globally confined to their initial location. The resulting structure is therefore a narrow disc.

In Figure \ref{fig8} the complex eccentricity map was (as expected) initially populated by particles clustered around the $e_{forced}$, since both eccentricity and alignment with the planet were forced (preventing the particles from gaining $e_{free}$). However, because particles do not share the same initial semi-major axis values ($45<a<80$~AU), particles have different $e_{forced}$ and precession times. Therefore the space occupied by particles expands from a cluster at $e=e_{forced}$ to form a filled circle with scatter and $0<e<0.3$ in the complex eccentricity map after 10 $t_{sec}$. 

\begin{figure*}
\begin{center}               
 \subfloat{\includegraphics[width=80mm,height=55mm]{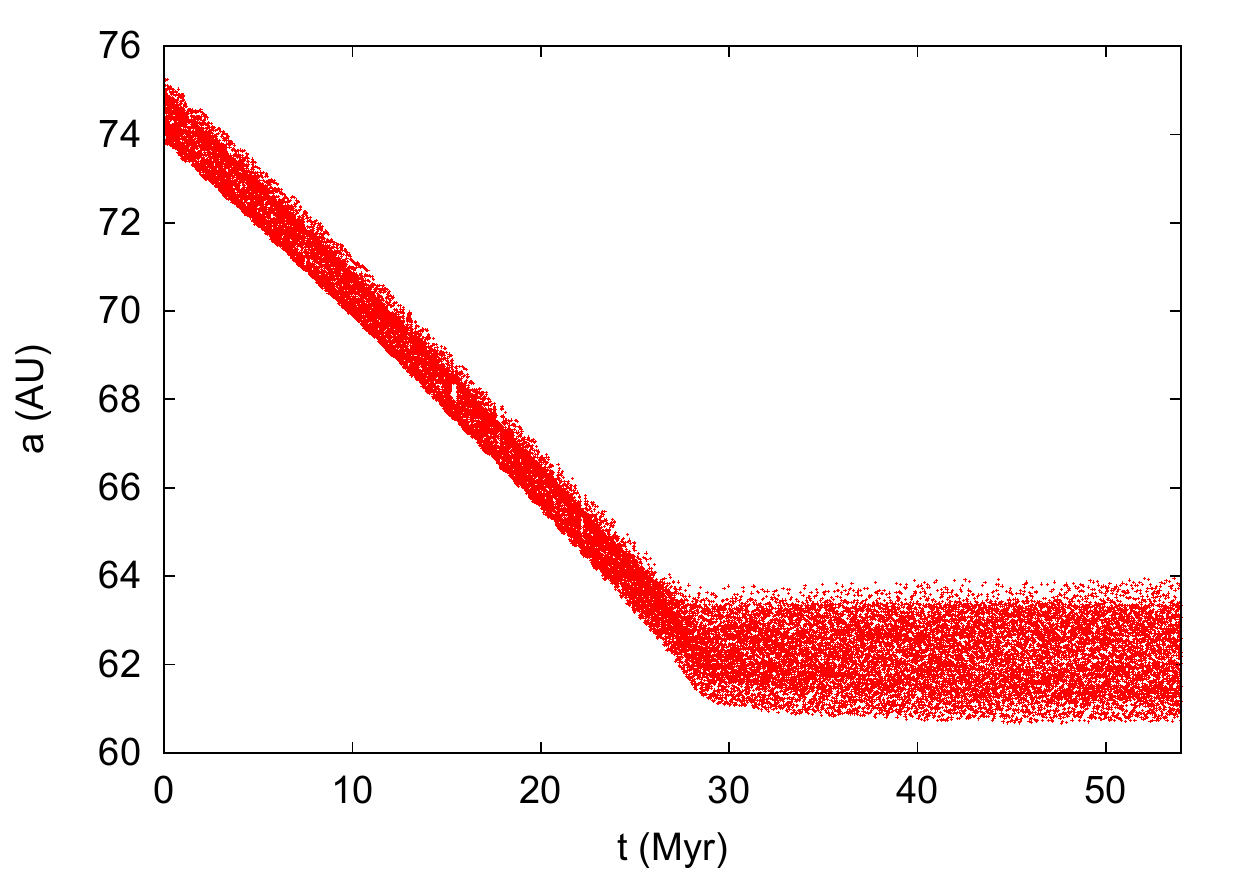}}
 \subfloat{\includegraphics[width=80mm,height=55mm]{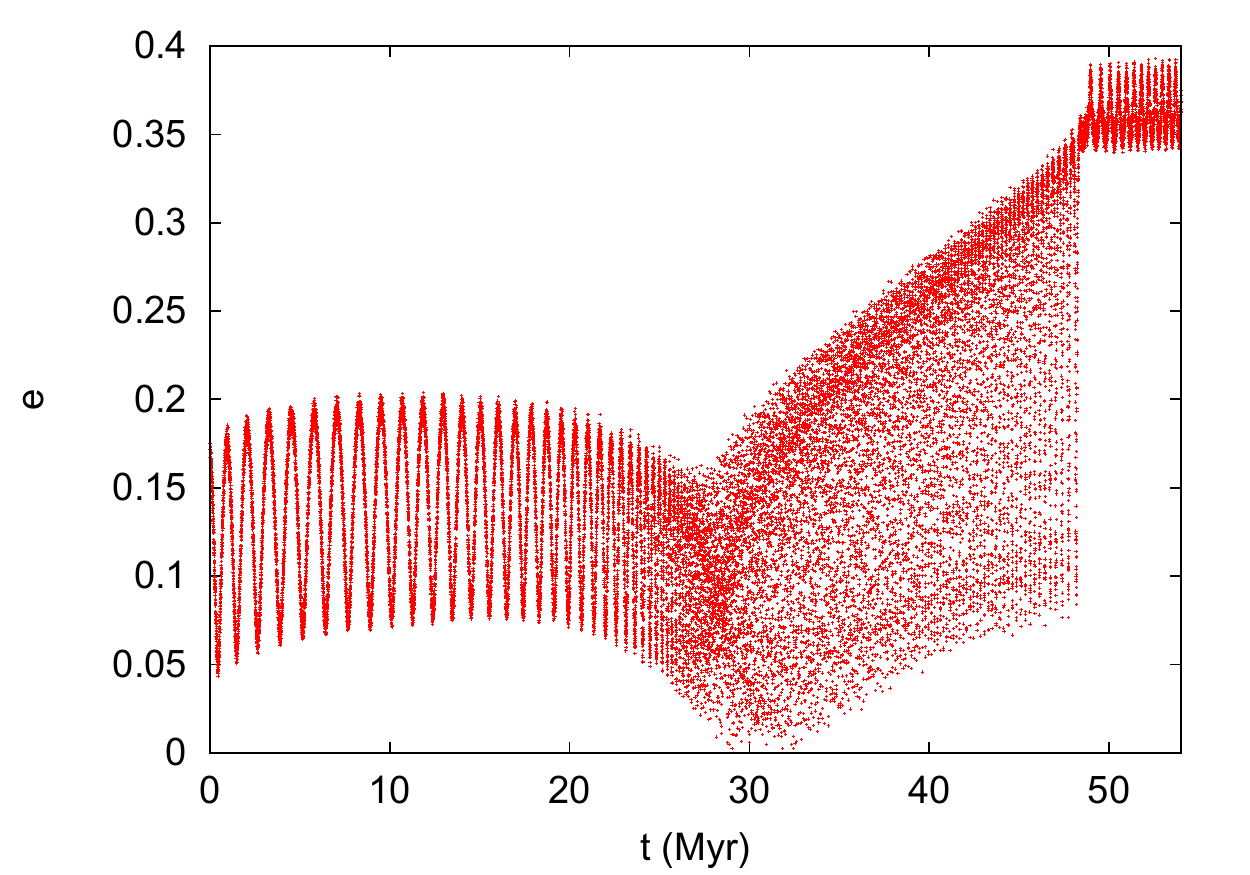}}
  \caption{Semi-major and eccentricity evolution for 20 $t_{sec}$ for a test particle which becomes trapped in the 3:1 MMR at $t=9.5~t_{sec}$ in simulations with a broad parent body belt from initial conditions of Class IIa (model 3).} 
\label{fig9}  
\end{center}
\end{figure*}

Contrary to models from the other classes, the occupancy profile shows the 3:1 MMR (located at 62 AU, near the center of the parent body belt) becoming more strongly populated with time. This is a natural consequence of the eccentricity and apse alignment being initially forced: although a differential precession rate induces some scatter in the particles' eccentricity, the initial forcing results in the maximal eccentricity acquired by the particles at $t=10~t_{sec}$ ($e < 0.3$) to remain lower than the critical eccentricity needed for a disruptive planet encounter to occur ($e \sim 0.4$). In addition to the ideal location of the 3:1 MMR at the center of the parent body belt, the particles moderate eccentricity allows them not only to stay trapped in the resonance for a longer time but also for particles from the outer parent body belt (65-80~AU) dragged in by the radiation forces to become trapped at later time as well, efficiently populating the resonance with time -- see Figure \ref{fig9}.
This effect has an impact on the width of the disc: while the final disc offset ($\delta=15$~AU), peak brightness location ($r_{0}= 48$~ AU), and disc eccentricity ($e_{disc}=0.3$) are similar than those of others classes within 1\%, the final disc width ratio is indeed narrower with $\Delta r/r_{0}= 0.3$ for this Class IIa compared to $\Delta r/r_{0} \sim 0.45$ for the other classes. 

In order to check if the disc width ratio value would move to 0.45 if particles were ejected from the 3:1 MMR, we ran the simulation until $t=20~t_{sec}$. We found that most particles remain trapped in the MMR and that little evolution occurred (see Figure \ref{fig9}).

\subsubsection{From a narrow parent body belt}
\begin{figure*}
\begin{center}               
 \subfloat{\includegraphics[width=67mm,height=48mm]{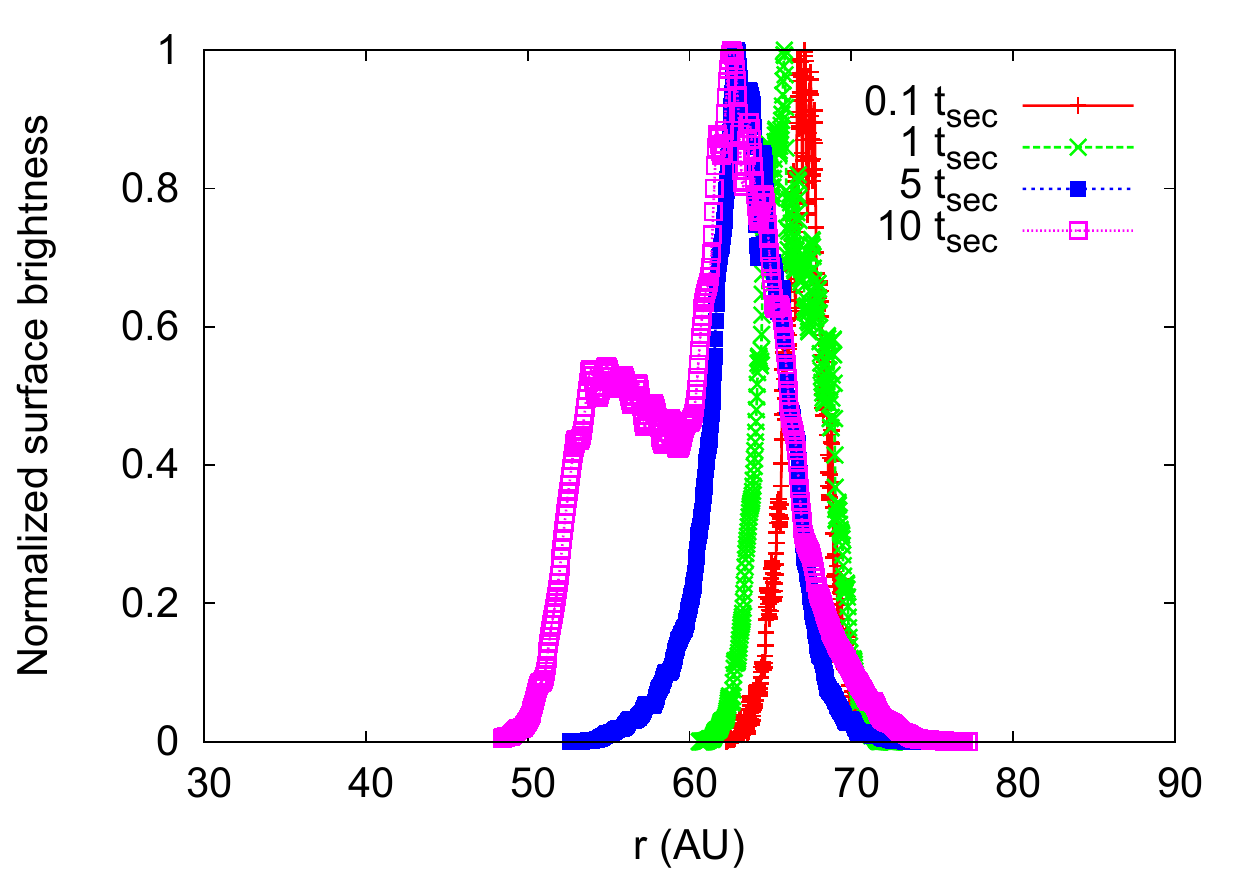}}  
 \subfloat{\includegraphics[trim=16mm 0cm 5mm 0mm,keepaspectratio=true,clip=true,width=70mm,height=48mm]{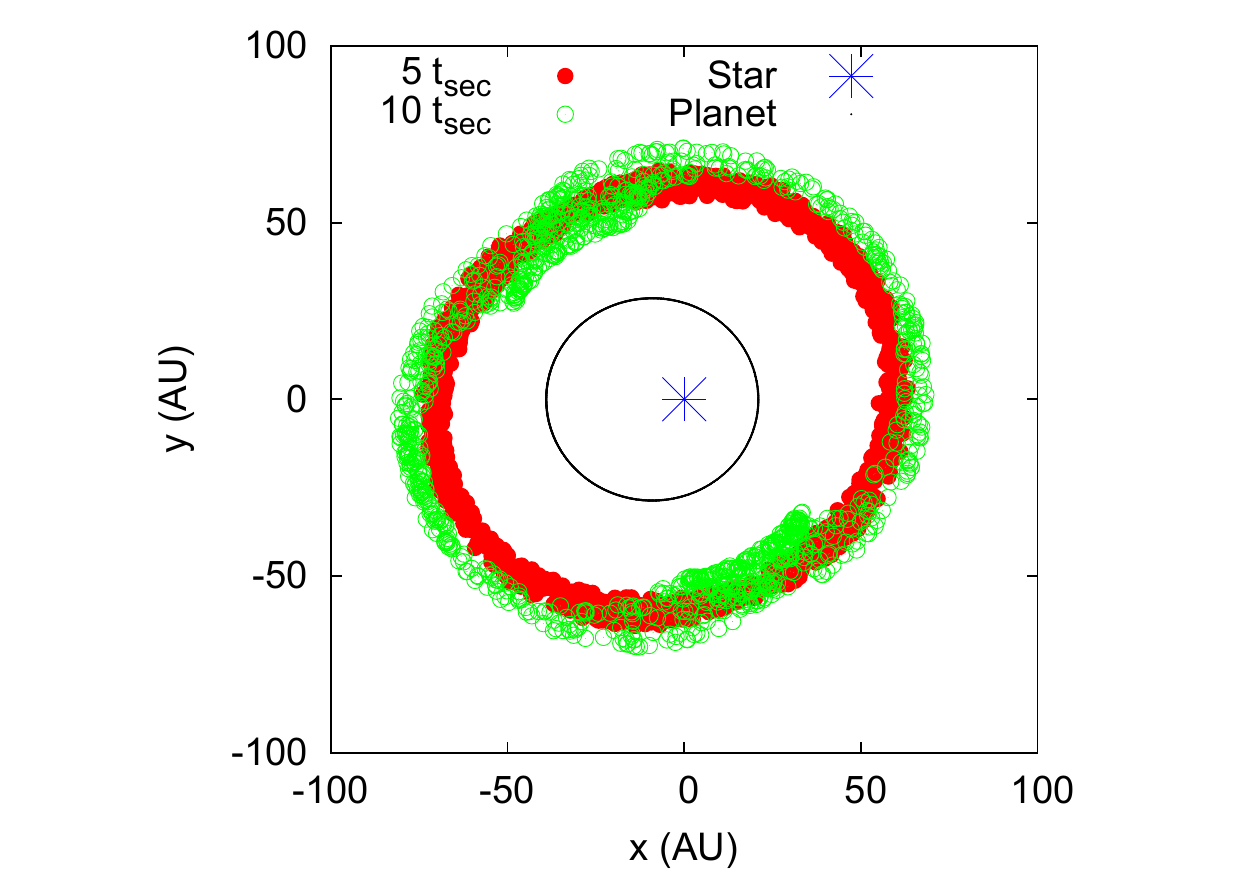}} 
 \subfloat{\includegraphics[trim=17mm 0cm 2mm 0mm,keepaspectratio=true,clip=true,width=70mm,height=48mm]{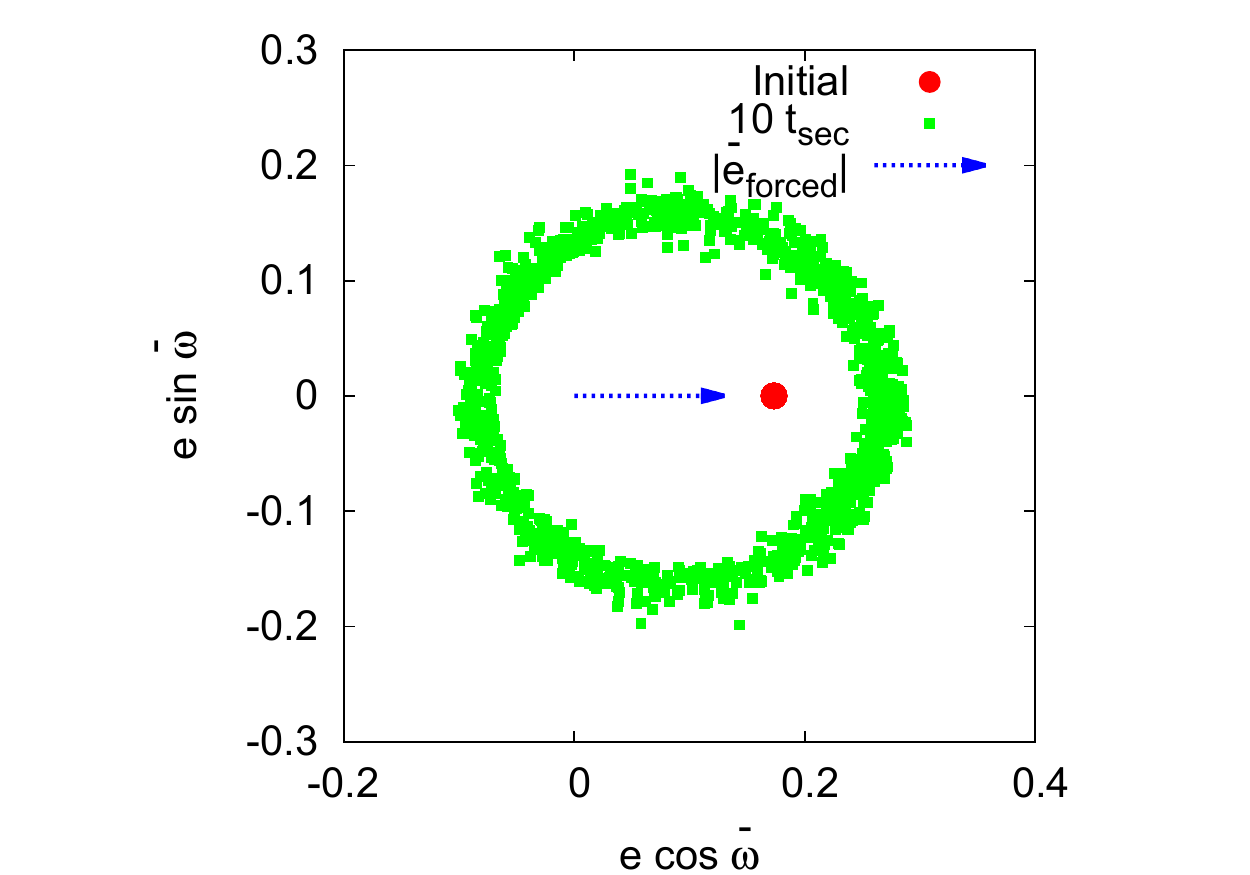}}
   \caption{Results for model 4, corresponding to a disc with an initially narrow parent body belt from initial conditions of Class IIa. Left : Normalised surface brightness profile at different epochs: 0.1, 1, 5 and 10 $t_{sec}$. Middle: particles distribution map enlighting the particles positions at $t=5~t_{sec}$ (red) and at $t=10~t_{sec}$ (green). Right: ($e \cos \overline{\omega}, e\sin \overline{\omega}$) complex eccentricity map.} 
\label{fig10}  
\end{center}
\end{figure*}

\begin{figure*}
\begin{center}               
 \subfloat{\includegraphics[width=80mm,height=55mm]{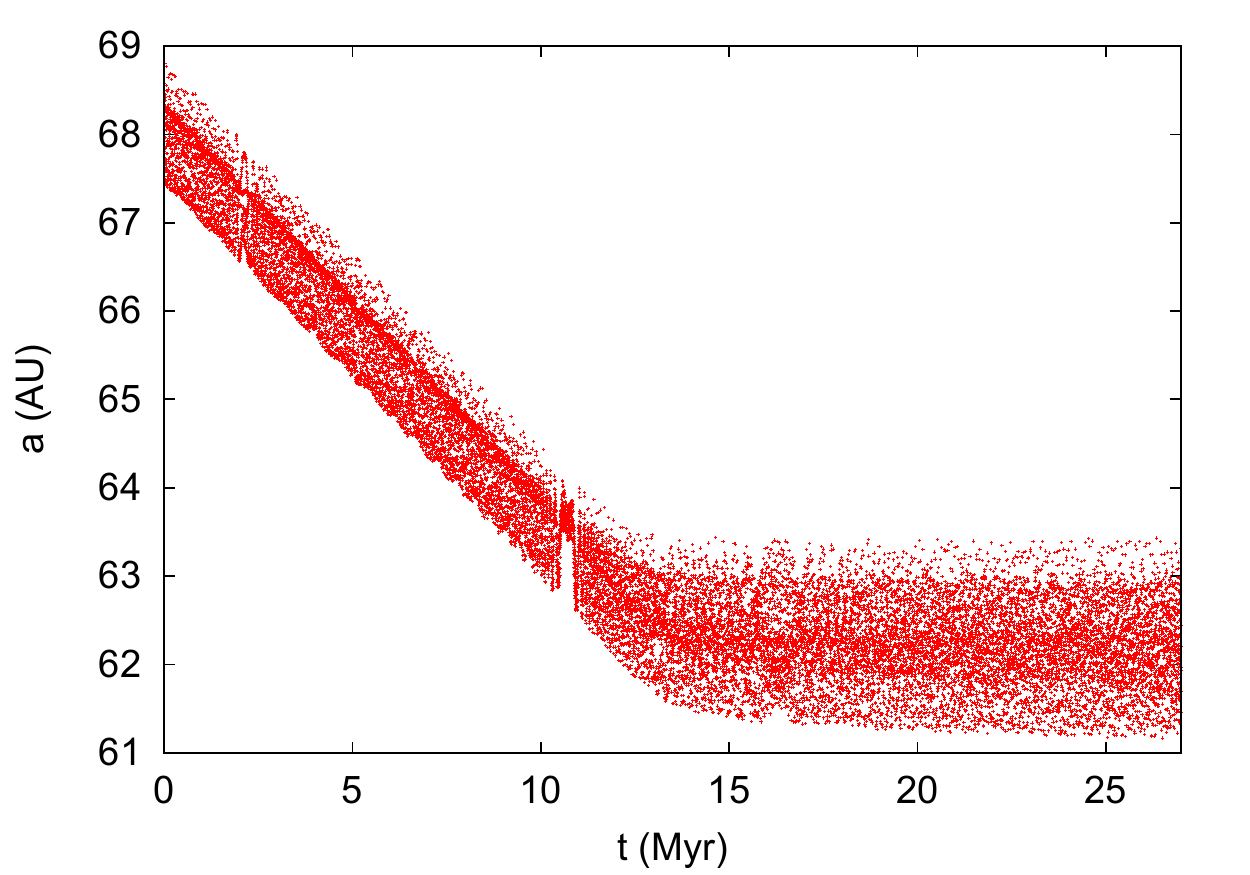}}
 \subfloat{\includegraphics[width=80mm,height=55mm]{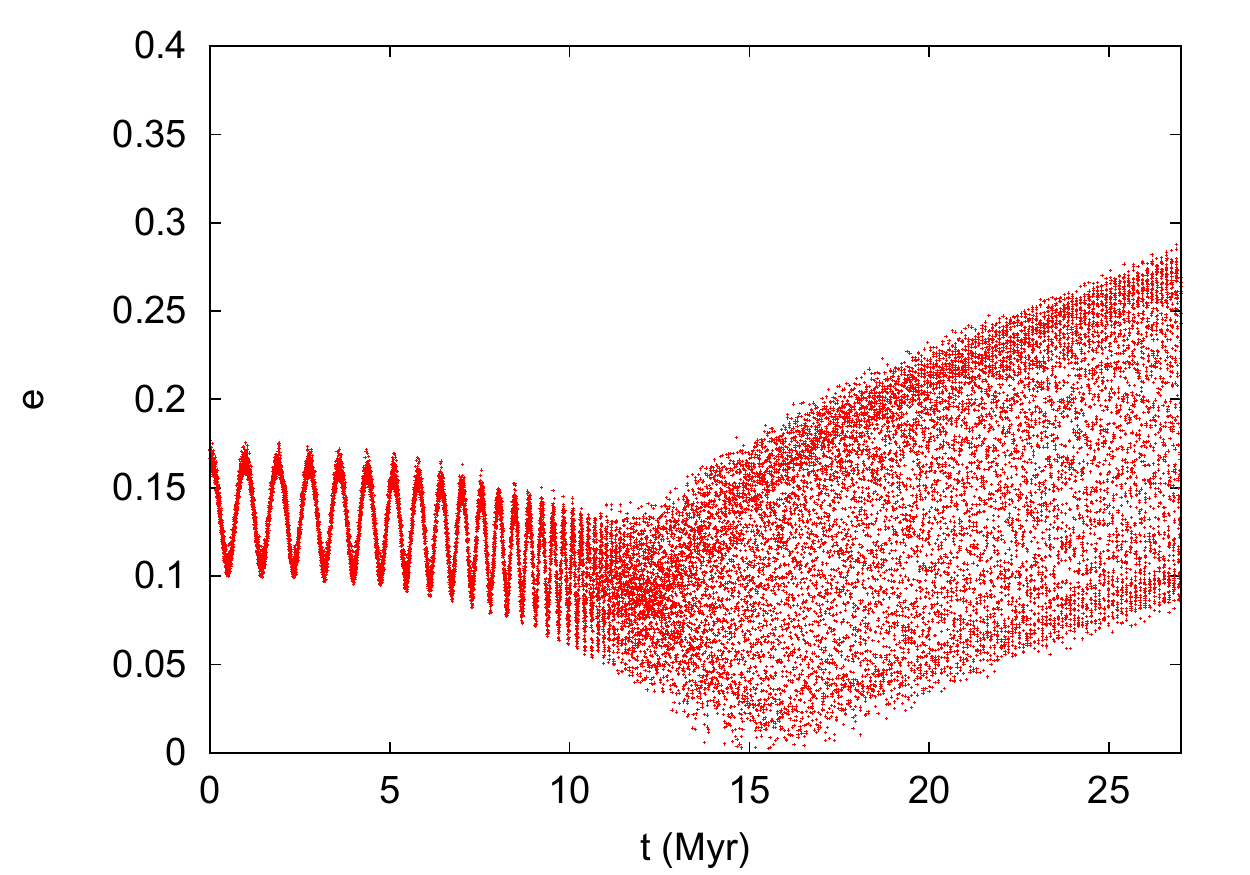}} \\
  \caption{Dynamical evolution of model 4. Semi-major (left) and eccentricity (right) evolution of the last surviving particle in the disc, representative of the evolution of the grain population in the disc.} 
\label{fig11}  
\end{center}
\end{figure*}

Only one of our models starting with an initially narrow parent body belt (model 4, Class IIa) resulted in a narrow disc - see Figure \ref{fig10} and Figure \ref{fig11}. This model is the extreme case of secular forcing where particles initially share a similar semi-major axis ($67.5<a<67.6$~AU) and will (i) initially maintain their forced eccentricity and remain confined until $5 ~t_{sec}$, (ii) get trapped in the nearest MMR around $5~t_{sec}$ until the end of the simulation at $10~t_{sec}$. The overall resulting disc structure has a peak brightness location located closer to the initial parent body belt and is narrower than model 3, the equivalent model of Class IIa with an initially broader parent body belt.  

To understand the disc dynamics, we look at the dynamical evolution of the semi-major axis and eccentricity of the last surviving particle in the disc in Figure \ref{fig11}: starting in the parent body belt located at $a=67.5$ AU, the grain is dragged inward by the radiation forces until it gets trapped by the 3:1 MMR with the planet at $t \sim 12\times 10^{6}$ years ($\sim$ 4.5 $t_{sec}$). As a result of this resonant trapping, its eccentricity is pumped up to reach $e \sim 0.25$ by $t=10~t_{sec}$.  In the complex eccentricity map (Figure \ref{fig10}), we see the test particles were initially confined at the forced eccentricity location, before forming a circle with $0 < e <0.25$ after gaining free eccentricity due to resonance trapping. This increase in particle eccentricity has consequences for the disc width, which can be seen in the distribution map in Figure \ref{fig10}. The initial thin ring is visible until $t=5~t_{sec}$ as particles start to be trapped in the MMR, and as eccentricity increases during the resonance, particles are seen to clump at two libration centers of the 3:1 MMR by $t=10~t_{sec}$. These clumps reach a diameter of 15~AU by the end of the simulation. The impact of the clump width can be seen in the normalized surface brightness profile in Figure \ref{fig10}: the ring has $\Delta r/r_{0} =0.08$ at 5 $t_{sec}$, before reaching $\Delta r/r_{0} =0.2$ after 10 $t_{sec}$.

In order to check if the disc clumps could grow enough for the surface brightness profile to recover a broader shape ($\Delta r/r_{0} \sim 0.26$) similar to the other classes of initial conditions, we ran an additional simulation with four times as many particle for a duration of 20 $t_{sec}$. However, after losing the first particle at $t=29\times10^{7}$ years ($\sim 11~t_{sec})$, we found the disc to be short lived, with the last particle is ejected at $t=32\times10^{7}$ years ($\sim 12.2~t_{sec}$). This results from resonance increasing particles' eccentricity to the critical value of $e > 0.3-0.35$, leading to a (dangerously) close encounter with the planet. As the disc width ratio shows little evolution between $10$ and $12.2~t_{sec}$ ($\Delta r/r_{0} =0.205$), we use the disc structure at $t=10~t_{sec}$ as the final disc structure with $\Delta r/r_{0}= 0.2$, $\delta=6.6$ AU and $r_{0}=62$ AU, leading to a disc eccentricity of $e_{disc}=0.11$.\\

Therefore we conclude that a disc with an initially secularly forced eccentricity and apse alignement with the planet leads to a narrower disc than the others classes of initial conditions. Again, it seems that the dynamics of a narrow parent body belt is singularly dominated by, in this case, the 3:1 MMR, while the broad parent body belt is shaped by resonance and secular interactions. In addition, the difference in the final disc structure between an initially narrow versus broad parent belt is strengthened when using an initial condition from Class IIa, with an apse aligned disc and a forced eccentricity. 

\begin{figure*}
\centering
\includegraphics[width=13cm,height=18.6cm]{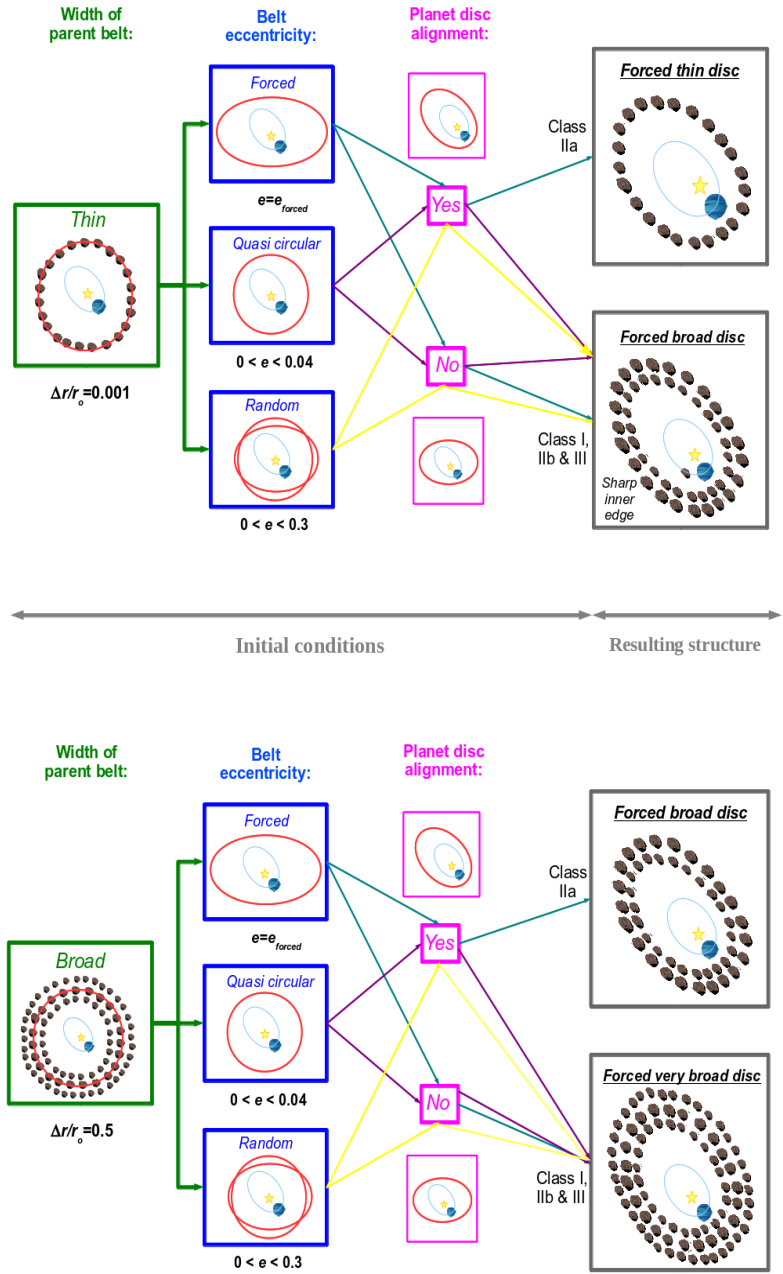} 
\caption{Summary of the possible initial conditions for a parent body belt interacting with a massive planet on an eccentric orbit, as well as the resulting disc structure for (top) an initially narrow parent body belt, and (bottom) an initially broad parent body belt, for varying initial belt eccentricities and planet-disc alignments.}
\label{fig12}
\end{figure*}

\begin{figure*}
\centering
\includegraphics[width=17cm,height=15cm]{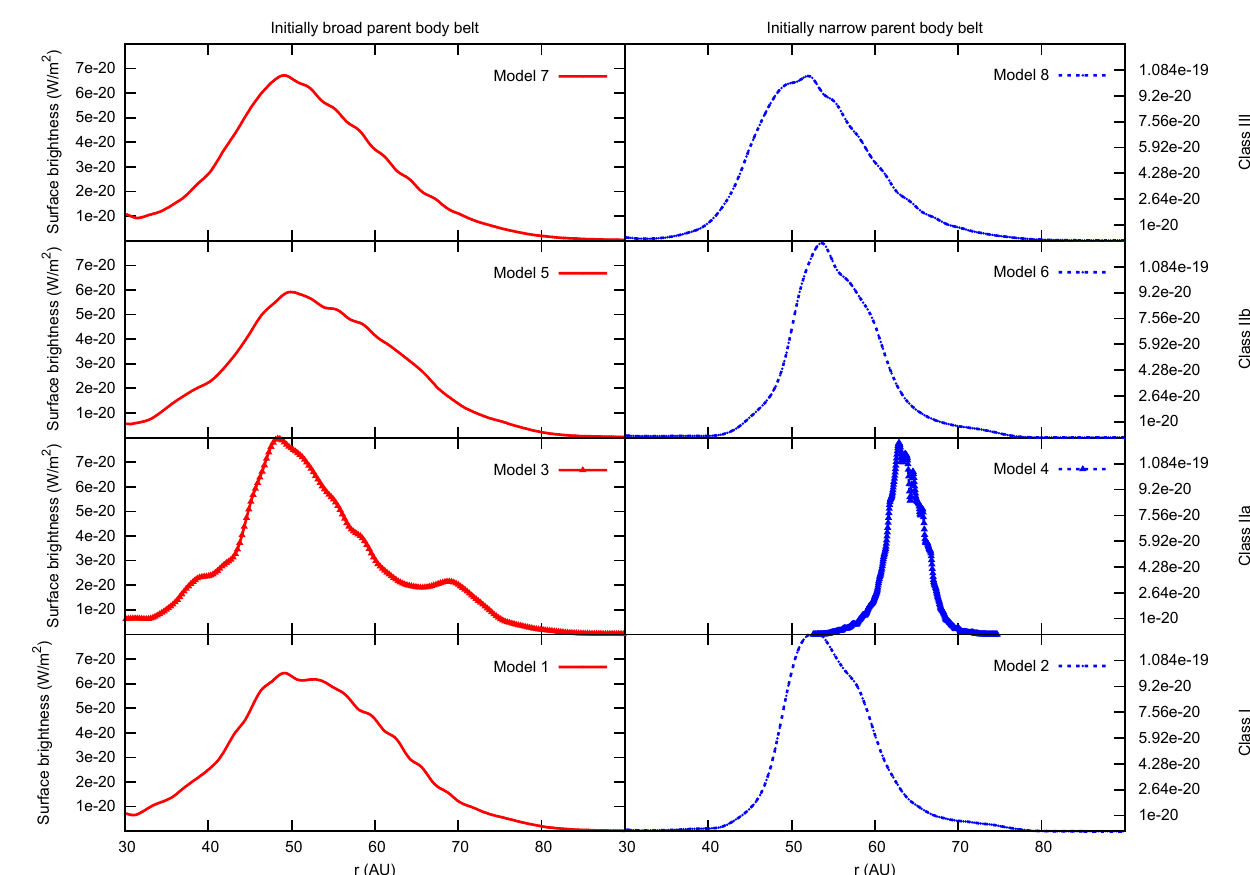} 
\caption{Surface brightness profiles at the end of the simulations for the 8 models. Similar profiles are observed for Class I, IIb and III for each category of parent body belt width, while the model from Class IIa always results in a narrower profil. Overall, models with a narrow parent body belt within each Class of initial conditions always result in narrower disc than discs with a broader parent body belt.}
\label{fig13}
\end{figure*}

A summary of all the possible outcomes from our models is illustrated in Figure \ref{fig12},while the final surface brightness profiles for all 8 models belonging to the different classes can be compared on Figure \ref{fig13}.

\section{Discussion \& Conclusions}
In this paper, we review the various initial conditions used in the literature to numerically model a debris disc interacting with a massive planet. We find that the initial conditions can be broadly divided into three classes: Class I, the dynamically cold disc; Class II, the secularly forced disc; and Class III, the dynamically warm disc. Their origins and when each should be used can be summarized as follows:
\begin{itemize}
\item The differences between Class I and III  are determined in the protoplanetary disc phase. Discs corresponding to Class I are usually used in simulations hosting a low mass planet (less than a few Jupiter masses) for which the eccentricity was damped by the gas disc before being excited by scattering or merging events. As a consequence, the disc remains quasi circular, and such initial conditions are good for debris disc systems harbouring low mass planet.
\item Systems with more massive companions (a few Jupiter masses) would have opened a gap in the gas disc, exciting the eccentricity of both the planet and disc, leading to a dynamically warm disc. Discs with massive companion should therefore be modelled using the initial conditions of Class III.
\item Class IIa and IIb, where the planet and disc are initially aligned, are based an analytical predictions from the secular perturbation theory. In both cases, the disc and planet are apse aligned, and in the Class IIa the disc eccentricity is set to the forced value by the the planet and thus the disc is expected to remain in this configuration. Class IIb, on the other hand, assumes the disc is initially quasi circular which is expected to gain eccentricity up to twice the forced value (since the free component is set to be equal the forced component). Class IIa discs could result from inelastic collisions damping the free eccentricity of the parent bodies and thus their eccentricity is set by a planet on eccentric orbit \citep{2006MNRAS.373.1245Q}.
\end{itemize}
We then run a suite of modified N-body simulations that model the interaction between a disc and a 2 Jupiter mass planet on an eccentric orbit interior to the parent body belt. We incorporate the radiation forces that act on the small grains of the discs. We explore 8 different initial conditions that cover all aspects of the three classes from the literature. We examine the consequences of varying the initial conditions on the resulting disc structure, as well as on the resonance and secular evolution of the disc. Our main results are:
\begin{itemize}
\item Models using initial conditions from Class I, Class IIb and Class III follow a similar evolution and result in similar disc structures. This is primarily caused by secular forcing of the eccentricity and the proximity of the planet removing particles on highly eccentric orbits, forcing all models to converge towards a similar structure. If the disc is not initially aligned, we find that the debris disc apse aligns with the planet within 1 $t_{sec}$, and its width increases after 0.5 $t_{sec}$ as a result of particle eccentricities gaining a free component and/or having different forced values precessing at  different timescales across the disc. 
\item Models using initial conditions from Class IIa result in a narrower disc than the other classes. This naturally arises from the eccentricity and apse alignment with the planet being forced, which prevents particles from acquiring free eccentricities and populating a broader disc. 
\item Discs with initially narrow parent body belts always result in narrower structures than discs with initially broad parent body belts, as the radiative drag on the dust in the inner region is delayed and trapping particles in outer MMRs is more difficult. Discs with an initially narrow parent body belt and initial condition from Class IIa represent the extreme case of secular forcing, where the dust remains strictly confined near the initial parent body belt. While very broad discs may be more accurately modelled by initially broad parent body belts, in the absence of collisions, narrower debris rings are expected to be best modelled by an initially narrow parent belt.
\item We stress, however, that modelling discs with initially narrow parent body belts is also more sensitive to initial conditions, because the confinement of particles makes the disc evolution strongly dominated by a single event, such as trapping by a particular resonance or eccentricity forcing, while initially broader discs have their evolution dictated by several events: multiple resonances, partial PR drag and secular eccentricity forcing.
\end{itemize}

To compare the outcomes of our numerical simulations with different initial conditions, we choose to study a particular and complex case of a debris disc interacting with a 2 Jupiter mass planet on a eccentric orbit, where the dynamics is dictated by a mix of secular and resonance interactions in addition to radiation forces. Running simulations with such planetary configuration combined with disc initial conditions of Class I and Class III seems in contradiction with the physical context we presented in Section 2. Although we choose to proceed in this way to model different dynamical interactions across a wide range of initial conditions to identify several outcomes, we also run additional simulations with all classes of initial conditions for all 8 models shown in Table \ref{Table2} using a 2 Jupiter mass planet but on a quasi circular orbit ($e_{p}=0.03$), where little secular forcing is expected to occur (i.e. Class IIa and Class IIb become a similar initial configuration with the disc initially apse aligned with the planet and respectively $e=e_{forced}=0.02$ and $0<e<0.04$). We find that (i) all models with an initially broad parent body belt from Class I, IIa, IIb and III resulted in a similar final disc structure, and (ii) the dynamics in discs with initially narrow parent body belts are purely shaped by resonances with a narrower structure than discs with an initially broader parent body belt.\\

Our aim is this study was to explore differences in the outcomes of numerical simulations with different initial conditions. We note however that additional physical interactions could contribute to the evolution of real debris discs systems. As pointed out by \cite{2015ApJ...798...83N}, grain-grain collisions is another factor which can play a key role in damping the effect of the initial parent body belt eccentricity in simulations.

\begin{acknowledgements}
This work was performed on the swinSTAR supercomputer at Swinburne University of Technology. E.T. is supported by a Swinburne University Postgraduate Research Award (SUPRA).  S.T.M. thanks the visiting professor scheme from University Claude Bernard Lyon 1 which partially supported this work. We thank Christophe Pinte for his help with \textit{MCFOST}, and the anonymous referee for helping to improve this paper.
\end{acknowledgements}

\bibliographystyle{plain}

\end{document}